\documentclass[a4paper,10pt]{article}

\usepackage{vmargin}
\usepackage{amsfonts}
\usepackage{amsmath}
\usepackage{graphics,graphicx,color}
\usepackage{psfrag}
\newcommand{\beq}{\begin{equation}}
\newcommand{\eeq}{\end{equation}}
\newcommand{\beqa}{\begin{eqnarray}}
\newcommand{\eeqa}{\end{eqnarray}}
\newcommand{\nn}{\nonumber}
\newcommand{\f}{\mathcal{F}}
\newcommand{\e}{\mathcal{E}}
\newcommand{\ve}{\mathcal{V}}
\newcommand{\W}{\mathcal{W}}
\newcommand{\veb}{{\mathcal{V}}_+}
\newcommand{\R}{\mathbb{R}}
\newcommand{\V}{\mathbb{V}}
\newcommand{\C}{\mathbb{C}}
\newcommand{\T}{\mathcal{T}}
\newcommand{\g}{\hat{G}}
\newcommand{\gh}{\hat{\mathfrak{g}}}
\newcommand{\G}{\mathcal{G}_{\T}}
\newcommand{\Vj}{\stackrel{j}{\mathbb{V}}}

\newcommand{\Pj}{\stackrel{j}{\pi}}
\newcommand{\h}{\hspace{1mm}}
\newcommand{\I}{\iota}

\def\vslash{\mbox{/ \hspace{-0.9em}$v$}}                

\setpapersize{USletter}
\setmarginsrb{39mm}{12mm}{28mm}{8mm}{12pt}{11mm}{0pt}{13mm}		     
\begin{document}

\title{\Large\bf Fermions in three-dimensional spinfoam quantum gravity}
\author{Winston J. Fairbairn\, \footnote{
fairbairn@cpt.univ-mrs.fr}\ 
\\[1mm]
\itshape{\normalsize{Centre de Physique
Th\'eorique \footnote{Unit\'e mixte de recherche (UMR 6207) du CNRS et des Universit\'es Aix-Marseille I, Aix-Marseille II et du Sud Toulon-Var. Laboratoire affili\'e \`a la FRUNAM (FR 2291).} }} \\ \itshape{\normalsize{Luminy - Case 907}} \\
\itshape{\normalsize{F-13288 Marseille, EU}}}
\date{{\small\today}}
\maketitle

\begin{abstract}
We study the coupling of massive fermions to the quantum mechanical dynamics of spacetime emerging from the spinfoam approach in three dimensions. 
We first recall the classical theory before constructing a spinfoam model of quantum gravity coupled to spinors. The technique used is based on a finite expansion in inverse fermion masses leading to the computation of the vacuum to vacuum transition amplitude of the theory. The path integral is derived as a sum over closed fermionic loops wrapping around the spinfoam. The effects of quantum torsion are realised as a modification of the intertwining operators assigned to the edges of the two-complex, in accordance with loop quantum gravity. The creation of non-trivial curvature is modelled by a modification of the pure gravity vertex amplitudes. 
The appendix contains a review of the geometrical and algebraic structures 
underlying the classical coupling of fermions to three dimensional gravity.  
\end{abstract}

\begin{flushleft}
CPT-P71-2006
\end{flushleft}

\section{Introduction}

The subtle interplay between matter and the classical geometry of spacetime is probably the deepest physical concept emerging from Einstein's equations. The becoming of this relationship in the Planckian regime, where the quantum mechanical aspects of spacetime become predominant, is a core issue in any attempt to the quantisation of gravity.
In this paper, we address the problem of fermionic couplings to three dimensional quantum gravity in the spinfoam approach. 

Spinfoam models \cite{SF} are discretized path integrals concretely implementing the Misner-Hawking three-geometry to three-geometry transition amplitude \cite{Misner}. They can be realised from different approaches. First, a spinfoam can be seen as a  spacetime history interpolating between two canonical states, i.e as a representation of the physical scalar product \cite{SFrep} of loop quantum gravity \cite{loop}. 
More generally, spinfoam models can also appear as a background independent, non-perturbative covariant quantisation of BF-like field theories by prescribing a regularization of the path integral of the theories combining (random) lattice field theory and group representation theory techniques in various dimensions. If the dimensionality of spacetime is different from three, where gravity is a BF theory, the local degrees of freedom of general relativity can by inserted directly into the discretized path integral by constraining BF theory down to gravity following Plebanski's prescription \cite{Pleb} directly at the quantum level \cite{BC}, \cite{BCd}. The path integral for gravity, called the Barrett-Crane model, is therefore obtained as a sum over a restricted subset of configurations. 
To obtain fully background independent models and to restore the infinite number of degrees of freedom of the theories, one works with a dual field theory formulation \cite{GFT1} generalising matrix models to higher dimensions. The idea is to write a field theory over a group manifold (GFT) whose Feynman amplitudes yield the spinfoam partition function defined on the cellular decompositions dual to the Feynman diagrams of the field theory. This approach has recently been rethought as a third quantised version of gravity, including a sum over topologies \cite{GFT2}.

In this paper, we explicitly construct a spinfoam model describing the coupling of massive fermionic fields to three dimensional Riemannian quantum gravity. We derive the spinfoam model from a classical action principle by giving a precise meaning to the vacuum to vacuum transition amplitude of the theory

\beq
\label{pathintegral}
\mathcal{Z}=\int \mathcal{D} e \hspace{.2mm} \mathcal{D} \omega \hspace{.2mm} \mathcal{D} \overline{\psi} \hspace{.2mm} \mathcal{D} \psi \hspace{.2mm} e^{i S[e,\omega, \overline{\psi}, \psi ]},
\eeq
\\
by regularising, in particular, the functional measures.
We proceed by expanding the fermionic determinant, obtained by integrating out the fermionic degrees of freedom, into closed fermionic loops.

The study of matter couplings to gravitation in the spinfoam approach has recently raised a large number of proposals (for models from the canonical perspective see \cite{mattercanonical}). Two classes of models seem to emerge from the overall picture: ideas converging towards unification, where matter is somehow already hidden in the models, and constructions where matter is added by coupling to the gravitational field.  
In \cite{Crane1}, a unified picture has been suggested where matter emerges as the low energy limit of the models derived from the configurations left out of the path integral for gravity when constraining BF theory down to gravity. The same author \cite{Crane2} proposes an alternative picture where matter appears as the conical singularities of the triangulated pseudo-manifolds arising in the GFT Feynman expansion. In \cite{Miko1}, the matter degrees of freedom are encoded in a subset of the finite dimensional irreducible representations of the symmetry group of gravity.
A supersymmetric three dimensional model based on the representation theory of $OSp(1 \mid 2)$ has been derived in \cite{Livine:2003hn}. In this approach, fermionic matter is ``hidden'' in the gauge field hence offering another unified picture. From the coupling perspective, different models for matter and different techniques have been implemented. Yang-Mills fields have been coupled to four dimensional quantum gravity in \cite{Miko2} and \cite{OP} using lattice field theory techniques where the lattice geometry is determined by the Barrett-Crane model. A spinfoam model for pure Yang-Mills has been derived in \cite{florian}.
Background independent point particle couplings have been introduced in three dimensional quantum gravity \cite{Freidel} (see \cite{Karim} for the canonical aspect) by using gauge fixing techniques or, equivalently, the de-Sousa Gerbert \cite{Sousa} algebraic description of the degrees of freedom of a point particle. 
GFT models have then followed \cite{GFT3} defining matter couplings to a third quantised version of gravity.
Recently, a subtle relationship between quantum field theory Feynman diagrams and three dimensional spinfoam models has been derived. Feynman diagrams have been shown \cite{barrett}, \cite{aristide} to yield natural observables of spinfoam models.
Also, in the no-gravity limit of spinfoam models coupled to point particles \cite{EL}, one recovers the Feynman diagram amplitudes of quantum field theory.
A generalisation of the above results to four dimensions has recently been proposed \cite{Baez:2006sa} from the canonical perspective. It is shown how the natural higher dimensional extension of three dimensional quantum gravity coupled to point particles leads to quantum BF theory coupled to strings and branes. The derivation of the dynamics of this proposal is currently under study \cite{W}. Finally, it has recently been realised \cite{artem} that the coupling of Wilson lines to the de-Sitter gauge theory formulation of gravity leads to the description of the dynamics of particles coupled to the gravitational field.

Here, we concentrate on the interplay between spinor fields and three dimensional quantum geometry.
The paper is organised as follows. Section two is a review of the classical coupling of fermions to the gravitational field. We recall, in particular, the effects of fermionic sources on spacetime torsion and curvature. The third section describes the construction of the model. We define the manifold and field discretizations, provide a simplicial action, define the fermionic loop expansion, and finally perform the integrations in the discretized path integral leading to a precise definition of the model. The appendix describes the geometrical and algebraic framework underlying the coupling of spinors to classical three-dimensional gravity.

\section{Classical theory}

In this first section, we derive the dynamical effects of fermions on the geometry of spacetime. All necessary definitions and conventions are given in the appendix. We first introduce the action governing the dynamics of the gravitational field before adding the fermionic term. We then derive the equations of motion of the coupled system.  

\subsection{Gravitational and fermionic action principles}

Consider a connected, oriented, compact, three dimensional differential manifold $M$ endowed with an Euclidean metric $g$ with diagonal form $\eta$ as our spacetime. We will denote $\mathcal{P}$ the bundle of $g$-orthonormal frames, that is, the principal bundle with base manifold $M$ and structure group $G=SO(\eta)=SO(3)$. Noting $V=\mathbb{R}^3$, we call $(\pi,V)$ the vectorial (fundamental) representation of $G$. Let $\hat{G}=Spin(\eta)=Spin(3)$ denote the spin group associated to the Euclidean metric $\eta$. We will choose as a basis of the real Lie algebra $\hat{\mathfrak{g}} = \mathfrak{spin}(3)$ (resp. $\mathfrak{g} = \mathfrak{so}(3)$) the set of generators $\{X_a\}_a$ (resp. $\{T_a\}_a$), $a=1,2,3$.
We will assume that the manifold $M$ is endowed with a given spin structure, i.e. a $\hat{G}$-principal bundle $\hat{\mathcal{P}}$ mapped with a two-to-one bundle homomorphism $\hat{\chi}$ onto the bundle of orthonormal frames $\mathcal{P}$. Note that the principal bundle $\mathcal{\hat{P}}$ is necessarily trivial. Let $\omega$ denote a metric (but not necessarily Riemannian) connection on $\mathcal{P}$ image under the Lie algebra isomorphism $\chi_* \equiv \pi_* : \mathfrak{spin}(3) \rightarrow \mathfrak{so}(3)$; $X_a \rightarrow T_a$, of a spin connection $\hat{\omega}$ on $\hat{\mathcal{P}}$. The data of the manifold $M$ together with the non-Riemannian metric connection $\omega$ defines the Riemann-Cartan structure underlying the framework necessary to model the interaction of fermions with the dynamical spacetime $M$.

We will use the fact that in the adjoint representation of the Lie algebra $\mathfrak{spin}(3)$ the matrix elements of the images of the generators are given by ${\pi}_*(X_a)^I_{\;\;J} = T_{a \hspace{1mm} J}^{\hspace{.5mm} I}=-\epsilon_{a \hspace{1mm} J}^{\hspace{.5mm} I}$, where the symbols $\epsilon_{a \hspace{1mm} J}^{\hspace{.5mm} I}$ denote the structure constants of the Lie algebra normalised such that $\varepsilon_{123}=-\epsilon_{132}=1$. We will furthermore use the isomorphism of vector spaces between $\mathfrak{spin}(3)$ and $V=\mathbb{R}^3$, regarded as the adjoint representation space of $\mathfrak{spin}(3)$, to make no distinction between the Lie algebra indices $a,b=1,2,3$ and the vector space indexes $I,J=1,2,3$.

The basic dynamical fields of Einstein-Cartan gravity are the soldering one-form $e=e^I \otimes e_I$ and the pull-back to $M$ by local trivialising sections of a principal metric connection on $\mathcal{P}$.
In a local chart $(U \subset M, x^{\mu}:U \rightarrow \mathbb{R}^3)$, the co-frame decomposes into a coordinate basis of the co-tangent space $e^I = e^I_{\mu} dx^{\mu}$, and the metric connection on $M$ is given by $\omega^{IJ} = \omega_{\mu}^{IJ} dx^{\mu} = \omega_{\mu}^a \otimes T_a^{IJ} dx^{\mu} \equiv -\epsilon^{IJ}_{\;\;K} \omega_{\mu}^K dx^{\mu}$.

Since we are focusing on the inclusion of fermionic matter, it will be convenient to work in the spinor representation of $2+1$ gravity (see e.g. \cite{Matschull}). To this aim, we define the linear map $\sigma : V \rightarrow End \h (\C^2)$, mapping the Clifford algebra $\mathcal{C}(3,0) \equiv \mathcal{C}(3)$ (in which the (dual of the) vector space $V$ is embedded) onto the Pauli algebra, i.e. the endomorphism algebra of the two-dimensional complex vector space $\V \equiv \C^2$ (see appendix). We consider as the dynamical fields of the theory the image ${\bf e}$ of the soldering form $e$ under the map $\sigma$ and (the pull-back to $M$ by a global trivialising section of) a principal connection on the spin bundle $\hat{\mathcal{P}}$, that is, a spin connection $\hat{\omega}$:

\beq
{\bf e} := \sigma(e) = e^I \sigma_I \hspace{3mm} \mbox{and} \hspace{3mm} \hat{\omega} = -\frac{1}{4} \omega^{IJ} \sigma_I \sigma_J =\frac{i}{2} \omega^I \sigma_I,
\eeq
\\
where the symbols $\sigma_I = \sigma(e_I)$ denote the Pauli matrices. 

Hence, we can think of the soldering form ${\bf e} : T_p M \rightarrow M_2(\C)$, forall $p$ in $M$, as a local isomorphism mapping any tangent vector into a two by two complex matrix, image in spin space of the associated vector in inertial space $V$: $\forall v \in T_pM, \h {\bf e}(v) = {\bf v} = v^I \sigma_I$. More precisely, the associated matrix ${\bf v}$ is traceless and Hermitian. Accordingly, we can think of the co-frame as taking value in $i \mathfrak{spin}(3) \simeq i \mathfrak{su}(2)$.

\newpage

The torsion $\hat{T} = \sigma (T^I e_I)$ and curvature $\hat{R}= - \frac{1}{4} R^{IJ} \sigma_I \sigma_J$ of the spin connection are then extracted from to torsion and curvature of the metric connection $\omega$:

\beqa
T^I &=& de^I + \epsilon^I_{\hspace{1mm} JK} \omega^J \wedge e^K \\ \nn
R^I &=& d\omega^I + \frac{1}{2} \epsilon^I_{\hspace{1mm} JK} \omega^J \wedge \omega^K ,
\eeqa
\\
where $\omega^I = - \frac{1}{2} \epsilon^I_{\;\;JK} \omega^{JK}$.
\paragraph{Gravitational action.}

The classical dynamics of the three dimensional spacetime $M$ can be expressed in terms of the Palatini action evaluated in the spinor representation of $\mathfrak{spin}(3)$

\beq
S_{GR}[{\bf e},\hat{\omega}]= \frac{i}{\kappa} \int_M  Tr \left({\bf e} \wedge \hat{R}[\hat{\omega}] \right),
\eeq
\\
where $Tr$ is the trace on $M_2(\C)$, and $\kappa$ is related to the Newton constant $G$ by $\kappa=8\pi G$. 

Equivalently, we can write the Palatini action in the vector representation of the spin group

\beq
S_{GR}[e,\omega] = \frac{1}{2 \kappa} \int_M \epsilon_{IJK} e^I \wedge R^{JK}[\omega] = -\frac{1}{ \kappa} \int_M  e^I \wedge R_I[\omega],
\eeq
\\
where we have used $Tr (\sigma_I \sigma_J \sigma_K) = 2i \epsilon_{IJK}$.

\paragraph{Fermionic action.}

The coupling of Dirac fermions $\psi \in C^{\infty}(M,\C) \otimes \V$ to $3d$ gravity is given by the following real action

\beq
\label{Da}
S_D[{\bf e}, \hat{\omega}, \overline{\psi}, \psi] = \frac{1}{2} \int_M \left[ \frac{1}{2} \left( \bar{\psi} \h {\bf e} \wedge {\bf e} \wedge \nabla \hspace{.3mm} \psi - \nabla \hspace{.3mm} \bar{\psi} \wedge {\bf e} \wedge {\bf e} \h \psi \right) + \frac{i}{6} m \h Tr ( {\bf e} \wedge {\bf e} \wedge {\bf e} ) \h \bar{\psi} \psi \right], 
\eeq
\\
where $m \in \mathbb{R}^+$ is the mass of the fermion field, the Dirac conjugate $\bar{\psi}$ is given by the Hermitian conjugate $\psi^{\dagger}$ (see the appendix) and the symbol $\nabla := \nabla(\hat{\omega})$ denotes the covariant derivative with respect to the spin connection. Equivalently, using the fact that the Clifford algebra $\mathcal{C}(3)$ is also a Lie algebra, namely $\sigma_I \sigma_J = i \epsilon_{IJ}^{\;\;K} \sigma_K + \eta_{IJ}$, the Dirac action can be reexpressed as

\beq
\label{Da1}
S_D[e, \omega, \overline{\psi}, \psi]=\frac{1}{2} \int_M \epsilon_{IJK} e^I \wedge e^J \wedge \left( \frac{i}{2} (\hspace{.3mm} \overline{\psi} \hspace{.3mm} \sigma(e^K)\hspace{.3mm} \nabla \hspace{.3mm} \psi  -  \nabla \hspace{.3mm} \overline{\psi} \hspace{.3mm} \sigma(e^K)\hspace{.3mm} \psi ) -\frac{1}{3} m e^K \hspace{.3mm} \overline{\psi}\hspace{.3mm}  \psi \right). 
\eeq
\\
The presence of the second term in the sum is to ensure the reality of the action keeping the same equations of motion for $\psi$ and $\overline{\psi}$. The introduction of such a term comes from the fact that in the presence of torsion, the usual Dirac vector current is not a total derivative \cite{torsion}.

\subsection{Equations of motion}

We now consider the equations of motion coming from the variation of the total action $S=S_{GR}+S_D$ with respect to the spin connection and to the (embedded) triad.

The variation of the connection yields the effects of the fermionic field on the spacetime torsion

\beq
\hat{T} = - \frac{\kappa}{4} \epsilon^I_{ \hspace{1mm} JK} \hspace{.3mm} \h \sigma_I \h e^J \wedge e^K \hspace{.5mm} \overline{\psi}\hspace{.3mm} \psi,
\eeq
\\
\newpage
while the variation with respect to the triad encodes the effects of matter on the curvature of spacetime :

\beq
\hat{R} = \frac{\kappa}{4} \hspace{.3mm} e^I \wedge \left( i ( \hspace{.3mm} \overline{\psi} \hspace{.3mm} \sigma^J \hspace{.3mm} \nabla \hspace{.3mm} \psi - \nabla \hspace{.3mm} \overline{\psi} \hspace{.3mm} \sigma^J \hspace{.3mm} \psi) - m \hspace{.3mm} e^J \hspace{.3mm} \overline{\psi} \hspace{.3mm} \psi \right) \h \sigma_I \sigma_J. 
\eeq
\\
From now on, we will set $\kappa =1$ and rescale the co-frame ${\bf e} \rightarrow  i {\bf e}$ such that it will be regarded as taking value in $\mathfrak{spin}(3) \simeq \mathfrak{su}(2) = \R \{ X_I \}_I$, with the convention $X_I = i \sigma_I$.
Having established the classical setting of the theory of Dirac fermions interacting with the geometry of an Einstein-Cartan spacetime, we now turn to the quantisation of the fermion/gravity coupled system.

\section{Spinfoam quantisation}

Feynman, following the intuition of Dirac, realised that quantum mechanical systems could be approached by two different kind of formulations: the canonical or Hamiltonian approach and the covariant or sum over paths formalism.
In the case of quantum gravity, the canonical approach is called loop quantum gravity and precisely predicts a discrete picture of quantum space. The fundamental excitations of the quantum gravitational field are one dimensional and supported by coloured, diffeomorphism invariant graphs. 
Such states are called spin networks. The world surface swept by a spin network 
forms a collection of two dimensional surfaces meeting on edges and vertices. The resulting object is a coloured two-complex called a spinfoam. A spinfoam can therefore be though of as a history of quantum space (a quantum spacetime) and a sum over spinfoams interpolating between two spin networks can be viewed as an implementation of the Misner-Hawking three-geometry to three-geometry transition amplitude for quantum gravity.

One of the most remarkable aspects of the spinfoam approach is the fact that a surprising number of independent research directions converge towards the same formalism. Indeed, spinfoams arise for instance, independently from loop quantum gravity, as a tool of the covariant formulation to compute the partition function of a large class of BF-like \footnote{By BF-like, we mean any theory expressable as a free topological BF part plus a polynomial function of the B
field \cite{GF}.} field theories. In the case of pure topological BF theory, the idea is to use the lack of local degrees of freedom of the theory to calculate the path integral with lattice regularization techniques without needing a continuum limit. In the case where local degrees of freedom are present, such an approach is therefore, in essence, only a (non exact) discretization of the theory of interest until a sum over triangulations is precisely understood. A fair amount of work has been devoted to such techniques, as recalled in the introduction.  

Three dimensional gravity is a topological field theory and hence nicely fits into the spinfoam framework. In fact, the first model of quantum gravity ever written was a spinfoam quantisation of riemannian three dimensional gravity: the Ponzano-Regge model \cite{PR}.
In this paper, we generalise this model to include the interaction of fermions with gravity by giving a concrete meaning to the sum over all possible paths in configuration space weighted by the complex exponent of the action, i.e the path integral or vacuum to vacuum transition amplitude.

In the presence of fermions, the theory has local degrees of  freedom.  Therefore the fixed triangulation that we will use below induces a physical cut-off of the degrees of  freedom of the theory.  In this paper, we do not address the problem of the removal of this regularization. We discuss some potential possibilities in the conclusion. 
  
This section is devoted to present the model. We first discretize the theory and regularise the path integral. We then integrate over the fermionic degrees of freedom and expand the resulting functional determinant in inverse fermion masses. This expansion yields a sum over fermion paths wrapping around the spinfoam. We finally compute the path integral of the coupled system.

\subsection{Discretization}
 
\subsubsection{Manifold discretization}

\begin{figure}[t]
\begin{center}
\psfrag{a}{$e_{1+}$}
\psfrag{b}{$e_{1-}$}
\psfrag{c}{$e_{2-}$}
\psfrag{d}{$e_{2+}$}
\includegraphics{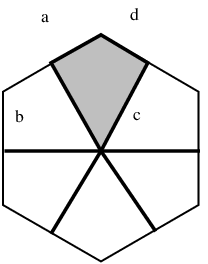}
\caption{A wedge.} 
\end{center}
\end{figure}

First, we pick a triangulation $\mathcal{T}$ of our spacetime manifold $M$ \footnote{Note that all compact three dimensional manifolds are triangulable.}. This means that we subdivide $M$ into a finite union of three-simplices (tetrahedra) $\Delta^3$ such that any pair of such simplices are either disjoint or meet on a common sub-simplex (triangle or segment). Two different three-simplices can have at most one common triangle. All simplices and sub-simplices are oriented and the orientations of two triangles in the boundary of two different three-simplices along which they are glued together are assumed to match; the one-simplices of those two triangles are assumed to have same coordinate length and must have pairwise opposite orientations.
If $n$ denotes the number of simplices of the triangulation $\T$, $M=\cup_{k=1}^n \Delta^3_k$. We will make the assumption that the simplicial manifold defined by $\T$ admits a spin structure. 
We furthermore assume that each tetrahedron $\Delta^3$, together with the four boundary tetrahedra glued along its four triangles, belongs to a given coordinate patch $U \subset M$ and that all three-simplices are diffeomorphic to the standard tetrahedron in $\mathbb{R}^3$.

We will work with the dual complex $\mathcal{T}^*$ of the triangulation or more precisely with its dual two-skeleton. This means that we place a vertex $v$ in the center of each tetrahedron $\Delta^3 \in \mathcal{T}$, link the vertices with edges $e$ going through the triangles $\Delta^2 \in \mathcal{T}$ and define a face $f$ as a closed sequence of edges intersecting the segments $\Delta^1 \in \mathcal{T}$.
We fix an orientation and a distinguished vertex for each face $f$ and call its edges $e_{1_f}, e_{2_f},...,e_{N_f}$ taken in cyclic order starting from the distinguished vertex.
The complex formed by the cells $(v,e,f)$ is the dual two-skeleton of $\mathcal{T}$. 

For consistency purposes that will become clear shortly, we furthermore refine the above cellular decomposition by introducing a subdivision of each dual face $f$ into so-called wedges $w$ (see \cite{Rei} and references therein). A wedge (see figure $1$) is defined by first identifying the center of a dual face $f$ as the point where the latter intersects the associated segment $\Delta^1$ of the triangulation. By drawing lines connecting the center of $f$ to the centers of the neighboring dual faces, one obtains a subdivision of $f$ into wedges $w$ consisting in the portions of the face lying between two such lines. In other words, a wedge belonging to a given face $f$ and to a given three-simplex $\Delta^3$ is the intersection between $f$ and $\Delta^3$, $w=f \cap \Delta^3$, as pictured in figure $1$. 
Each wedge acquires an orientation induced by the orientation of the face to which it belongs and is unambiguously associated to the simplex to which it belongs.
The complex formed by these finer $2$-cells, their boundary edges and the vertices lying at the endpoints of the boundary edges is called the derived complex $\mathcal{T}^+$. There are two types of $1$-cells in $\mathcal{T}^+$. We distinguish between the edges of the derived complex $\mathcal{T}^+$ whose intersection with the edges of the dual complex $\mathcal{T}^*$ is non empty and the others. We will note $e_+$ the edges of $\mathcal{T}^+$ belonging to $\partial w \cap \partial f$ and $e_-$ the edges exclusively in $\partial w$. Note that the $e_+$ type edges are exactly half edges of $\mathcal{T}^*$. There are three types of vertices around each wedge. The center of the face to which the wedge belongs, noted $c$, one vertex $v$ belonging to $\mathcal{T}^*$ and two vertices forming the intersection of the edges emerging from $v$ with the two triangles meeting on the segment defining the center of the associated face. These two new types of vertices will be noted $v_+$.  
 
We denote by $\mathcal{V},\mathcal{E},\mathcal{F}$ the set of vertices, edges and faces of $\mathcal{T}^*$ while the set of vertices $\ve \cup \{v_+\} \cup \{c\}$, edges $\{ e_- \} \cup \{ e_+ \} = \e_- \cup \e_+$ and wedges $\{ w \}$ of $\mathcal{T}^+$ will respectively be noted $\mathcal{V}_+$, $\mathcal{E}_{\pm}$ and $\mathcal{W}$. We will call $\mathcal{W}_v$ the set of wedges restricted to the wedges belonging to the tetrahedron dual to the vertex $v$.  
We will call $n(\ve) \equiv n$ the cardinality of the set $\ve$.

We then assume that the global trivialising section of $\hat{\mathcal{P}}$ is piecewise constant in each tetrahedron. Hence, the spinor and vector bundles associated to $\hat{\mathcal{P}}$ are also globally trivialised by a piecewise constant section.
We therefore have a basis $(e_I(v))_I$ of $V$ (or equivalently a basis $(X_I(v))_I$ of $\mathfrak{spin}(3)$) and a basis $(f_{\alpha}(v))_{\alpha}$ of $\mathbb{V}$ for each tetrahedron inside which the vector and spinor fields are globally defined.

\subsubsection{Simplicial field configurations}

We now turn towards the discretization of the field content of the theory. The main hypothesis is that the cellular decomposition is sufficiently refined so that the fields of the theory can be though of as piecewise constant inside each three-simplex $\Delta^3$.
Let the latin indices from the middle of the alphabet $i,j,...$ and from the
end of the alphabet $...,t,u,v$ respectively denote the vertices of the triangulation $\mathcal{T}$ and those of $\mathcal{T}^+$. We denote by $ab$ the segment linking the vertices $a$ and $b$. For each vertex $v \in \ve$,

\paragraph{A simplicial soldering form configuration} ${\bf e}$ is a map
\footnote{More precisely, a simplicial triad is a functor from the manifold groupoid $\pi(M)$ (the category whose objects are the vertices of the triangulation and the invertible morphisms are the edges linking the points) to the category $\mathfrak{spin}(3)$, where there is one object and the morphisms are the additive group operations.} 
\beqa
{\bf e} : \mathcal{W}_v   &\rightarrow& \mathfrak{spin}(3) \\ \nn
          w               &\rightarrow&  {\bf e}_w   .
\eeqa
\\
Here, ${\bf e}_{w}:={\bf e}_{ij}$ denotes the line integral of ${\bf e}$ along the segment $ij \in \T$ dual to the face containing the wedge $w$; ${\bf e}_{ij}={\bf e}(v_{ij})={\bf v}_{ij}=\int_{ij} {\bf e}=\int_J s_{ij}^* {\bf e}$, where $s_{ij} :J \subset \mathbb{R} \rightarrow M$ is the segment linking $i$ and $j$ and $v_{ij} \equiv \dot{s}_{ij}$ denotes the tangent vector. The components of ${\bf e}_{w}$ are measured with respect to the frame $(X_I(v))_I$ associated to the simplex containing the wedge $w$: ${\bf e}_w := e_w^I \h X_I(v)$. 
This is one of the reasons why the complex $\T^*$ needed to be refined to $\T^+$: a face $f$ being shared by different simplices and the trivialisation of $\hat{\mathcal{P}}$ being chosen to be piecewise constant in each three-simplex, the assignment of a simplicial triad configuration to the face $f$ would have been meaningless: with respect to which frame would we express it ? 

The wedges of the face $f$ are in one to one correspondence with the three-simplices neighboring the tetrahedron dual to the vertex $v$ and all share the same segment $ij$ as it defines the center of the face. Hence, the vectors in $\mathfrak{spin}(3) \simeq \mathbb{R}^3$ assigned to each of these wedges will in fact be the same vector measured with respect to the frames associated to the simplices containing each wedge. As we will see, the quantum treatment of the theory will naturally impose on this vector to have same norm in different frames.  

By using the coordinates $x^{\mu}$ associated to the patch $U \supset v$ , we see that, at first order in the segments lengths, we have $e_{ij}^I \simeq e_{\mu}^I(x_v) \Delta_{ij}^{\mu}$ where $x_v$ is the coordinate of the dual vertex $v$ (it could be any point inside $v^*$) and $\Delta_{ij}^{\mu} \equiv v_{ij}^{\mu}=x^{\mu}(j) - x^{\mu}(i)$ is the coordinate length vector of the segment $ij$.

\paragraph{The spin connection} $\hat{\omega}$ assigns a spin rotation to the edges belonging to $\mathcal{E}_+$, i.e. to the dual half edges (forming the intersection between the boundaries of the faces and the boundaries of the wedges)

\beqa
\omega : \mathcal{E}_+   &\rightarrow& Spin(3) \\ \nn
           e             &\rightarrow&  U_{e_+} .
\eeqa
\\
The spin rotation is defined to be the parallel transport matrix $U_{e_+}:= U_{uv}(g) = \overset{1/2}{\pi}_{uv}(g) = P \h exp^{\int_{uv} \hat{\omega}}, \h g \in \hat{G}$, of the spin connection. We have used the notation $\stackrel{j}{\pi} \h : \hat{G} \rightarrow Aut \h (\Vj)$ for the spin $j$ representation (see the appendix), and $P$ means path ordering. If we reverse the orientation of an edge, the spin matrix gets mapped to its hermitian conjugate : $U_{vu} = U_{uv}^{\dagger}$.

To complete the picture, we also assign group elements $U_{e_-}$ to the edges $\mathcal{E}_-$ of $\mathcal{T}^+$ whose intersection with the boundaries of the faces are empty, i.e the edges converging toward the centers $c$ of the faces. These holonomies will be called auxiliary variables as they carry only information about the topology of the faces to which the fermions will turn out to be insensitive. 
 
By using the coordinates associated to the patch $U \subset M$ containing the simplex dual to the vertex $v$, we see that, at first order in the edges lengths, the holonomy approximates the connection: $U_{uv} \simeq 1\!\!1 + \Delta_{uv}^{\mu} \hat{\omega}_{\mu}(x_v)$.  

The discretized curvature is assigned to the wedges and defined to be the holonomy $U_w$ of the spin connection along the boundaries of the wedges $w$ :

\beq
U_w = \prod_{e \in \partial w} U_e.
\eeq
\\
This is valid since the curvature is the obstruction to the closing of the horizontal lift of an infinitesimal loop. 
More precisely, $U_w$ is of the form $U_w = U_{e_{1+}} U_{e_{1-}} U_{e_{2-}} U_{e_{2+}}$ (see figure $1$).
The base point of the holonomy is fixed by discretized gauge invariance arguments to be the dual vertex $v \in \mathcal{T}^*$ \cite{GF}.

We now turn towards the simplicial field content of the fermionic action. Fermions, as sections of the spinor bundle, must live on $0$-cells of the discretized manifold.
The structure of the classical Dirac action \eqref{Da} furthermore shows us what type of $0$-cells should the simplicial fields be assigned to. The volume, or mass, term tells us that the fermions live on $0$-cells of $\T^*$ or $\T^+$ topologically dual to three dimensional regions and the area, or kinetic, term implies that the fermions travel on $1$-cells dual to objects supporting areas. By inspection, fermions can only be assigned to vertices of $\T^*$. Hence,

\paragraph{The fermionic fields} assign a spinor $\psi_v$  $\in \mathbb{V}$
(resp. a co-spinor $\overline{\psi}_v$ $\in \overline{\mathbb{V}}^*$) to each vertex $v$ of $\ve$,
\beqa
\psi, \hspace{1.5mm} (\overline{\psi}) : \ve &\rightarrow& \mathbb{V} \hspace{1.5mm} (\overline{\mathbb{V}}^*) \\ \nn
                            v                 &\rightarrow&  \psi_v  
\hspace{1.5mm} (\overline{\psi}_v).	    
\eeqa
\\
Our simplicial fermions live in the discrete analogue of the vector space \footnote{In fact, the space of smooth sections of the spinor bundle $\Gamma(SM)$ is a module over the commutative ring of smooth complex valued functions $C^{\infty}(M,\C)$.}
of sections $\Gamma(SM) \simeq C^{\infty}(M,\C) \otimes \V$ of the spinor bundle $SM$. This space is the vector space of $\V$-valued sequences $\{\psi \h \mid \h \psi : \ve \rightarrow \V \} \simeq \mathcal{S}(\ve) \otimes \V$ mapping the space of vertices $\ve \hookrightarrow \mathbb{N}$ into the Clifford module $\V$. Here, we have noted $\mathcal{S}(\ve)$ the vector space of complex valued sequences. We choose as a basis for $\mathcal{S}(\ve)$ the set $\{e^v\}_{v}$ of $n$ vectors in $\mathcal{S}(\ve)$ such that $e^v(u)=\delta^v_{u}$. Picking a basis $\{f_{\alpha}\}_{\alpha}$ in $\V$, a simplicial spinor field $\psi \in \mathcal{S}(\ve) \otimes \V$ reads $\psi= \psi_v^{\alpha} e^v \otimes f_{\alpha}$, with $\psi(v)=\psi_v^{\alpha} f_{\alpha} \in \V$, $\psi_v^{\alpha} \in \C$. We will note $\mathbb{S}$ the vector space $\mathcal{S}(\ve) \otimes \V$ of simplicial spinor fields.

We furthermore need to require that our simplicial fermions obey the the Fermi-Dirac spin-statistics. For instance, the fermionic {\itshape components}, i.e the coefficients $\psi_v^{\alpha}$ appearing in the expression of a simplicial spinor $\psi$ in our chosen basis must anticommute; $\psi_v^{\alpha} \psi_v^{\beta} + \psi_v^{\beta} \psi_v^{\alpha}=0$. 
To mathematically capture this physical fact, it is necessary to change the field of numbers on which the vector space $\mathbb{S}$ is built on from the field of complex numbers $\mathbb{C}$ to a supercommutative ring. This ring is given by a Grassmann algebra that we will note $\mathcal{G}_{\T}$ and which is constructed as follows \cite{simplicial spinors}.

Let $\V_v$ (resp $\overline{\V}^*_v$) denote the copy of the Clifford module $\V$ (resp of the dual complex conjugate Clifford module $\overline{\V}^*$) associated to the vertex $v$ of the dual complex $\T^*$. 

\newpage

Consider the complex vector spaces $E$ and $\overline{E}^*$ obtained as the direct sums   

\beq
E = \bigoplus_{v \in \ve} \V_v, \hspace{3mm} \overline{E}^* = \bigoplus_{v \in \ve} \overline{\V}^*_v.
\eeq
\\  
The Grassmann algebra $\mathcal{G}_{\T}$ is obtained as the exterior algebra over the complexified realification of $E$ (identifying $E$ with its dual vector space $E^*$):

\beq
\mathcal{G}_{\T} = \bigwedge \left( E \oplus \overline{E}^* \right).
\eeq
\\
As a complex vector space, $\mathcal{G}_{\T}$ is of dimension $2^{4n}$.
We finally make a chart choice on $\mathcal{G}_{\T}$ such that the algebra is defined by the generating system $\{\psi_v^{\alpha}, \overline{\psi}_{v\alpha}\}_{v,\alpha}$ of the (odd) elements of $\mathcal{G}_{\T}$ consisting in the components of the simplicial spinors and co-spinors.

Fermions will therefore not live in $\mathbb{S}$ regarded as a vector space but rather in $\mathbb{S}$ considered as a supervector space \cite{dewitt}, i.e a module (in fact, a bimodule) on the ring given by $\mathcal{G}_{\T}$. An element $\psi$ in $\mathbb{S}$ will be written as $\psi=\psi^{\alpha}_v e^v \otimes f_{\alpha}$ with $\psi^{\alpha}_v \in \Lambda^1(E \oplus \{0\}) \subset \mathcal{G}_{\T}$, where the former is included in the later as a vector subspace. Note that we now need to consider $\mathcal{S}(\ve)$ entering the definition of $\mathbb{S}$ as the module of Grassmannian valued sequences.

\subsubsection{Simplicial action}

We can now construct the discretized version of the fermion/gravity action. As a simplicial action, we take the real valued functional $S_{\T}=S_{GR,\T} + S_{D,\T}$, 
where $S_{GR,\T}$ and $S_{D,\T}$ are defined by
\beq
\label{GRaction}
S_{GR,\T}[{\bf e}_w, g_e]=\sum_w Tr \left( {\bf e}_w U_w \right) ,
\eeq
and
\beq
\label{Daction}
S_{D,\T}[{\bf e}_w, g_e, \overline{\psi}_v, \psi_v]= \frac{1}{2} \left( -\frac{1}{4} \sum_{uv} \left(
\overline{\psi}_u \hspace{.5mm} A_{uv} U_{uv} \hspace{.5mm} \psi_v - \overline{\psi}_v 
\hspace{.5mm} U^{\dagger}_{uv} \hspace{.5mm} A_{uv} \hspace{.5mm} \psi_u \right) - \frac{1}{3} \sum_v m V_v \overline{\psi}_v \psi_v \right), 
\eeq
\\
where $e \in \mathcal{E}_{\pm}$ (resp. $e \in \mathcal{E}$) for the gravitational part (resp. the fermionic part) and the sums in the simplicial gravitational and Dirac actions are respectively over all wedges and all edges $uv$ linking the adjacent vertices $u$,$v$.
In equation \eqref{GRaction}, the trace is taken in fundamental representation of the spin group. In \eqref{Daction}, 
\beqa
\label{area}
A_{uv} &\equiv& A_{uv}({\bf e}_w) = \frac{1}{3} \sum_{w,w' \supset uv} {\bf e}_w {\bf e}_{w'} sgn(w,w') \\ \nn
       & = & - A_{I \hspace{.5mm} uv} i \sigma^I ,  
\eeqa
\\
with 
\beq       
A_{I \hspace{.5mm} uv} = \frac{1}{3} \sum_{w,w' \supset uv} \epsilon_{IJK} e^J_w e^K_{w'} sgn(w,w'),
\eeq
\\
is the discretized version of the two-form ${\bf e} \wedge {\bf e}$ appearing in the kinetic term of the fermionic action. This term is the spin space analogue of the two-form $(* e \wedge e)$ where the star $*$ denotes the hodge operator acting on the exterior algebra $\Lambda(V)$ over $V=\mathbb{R}^3$. At the simplicial level, we are looking at the integrated version of this two-form on the triangles of $\mathcal{T}$ or, equivalently, we are considering the
evaluation of this two form on all possible couples of vectors $(v_{ij},v_{ik})$ generating each infinitesimal triangle $ijk$ of the triangulation. On a particular triangle $(v_{ij},v_{ik})$, $({\bf e} \wedge {\bf e})(v_{ij},v_{ik}) = -2i \sigma^I (* e \wedge e)_I (v_{ij},v_{ik})$, with

\beq
(* e \wedge e)_I (v_{ij},v_{ik})= \frac{1}{2} \epsilon_{IJK} e^J \wedge e^K (v_{ij},v_{ik}) = \frac{1}{2} \epsilon_{IJK} (v_{ij} \wedge v_{ik})^{JK} = (v_{ij} \times v_{ik})_I ,
\eeq
\\
where $(u \wedge v)^{IJ}:= 2 u^{[I} v^{J]}$ and $\times$ denotes the usual cross product on $\mathbb{R}^3$. This evaluation therefore yields (the image in $\mathfrak{spin}(3)$ of) the internal normal area one-form to the triangle $ijk$ measured in the inertial frame $(e_I(x_v))_I$ associated to the simplex dual to $v$ which is one of the two tetrahedra containing the triangle. 
In expression \eqref{area}, the sum is taken over the three possible pairs (without counting the permutations) of wedges meeting on the edge $e=uv$. This explains the $1/3$ factor. Finally, $sgn(w,w')$ equals $\pm 1$ depending on the sign of the (coordinate) area bivector. This factor ensures the antisymmetry of forms at the simplicial level. 
     
The $V_v$ symbol assigned to each dual vertex denotes another polynomial function of the discretized gravitational field which is given by

\beqa
\label{volume}
V_v &\equiv& V_v({\bf e}_w) = \frac{1}{2(16 \times 3!)} \sum_{w,w',w'' \supset v} Tr \h ( {\bf e}_w {\bf e}_{w'} {\bf e}_{w''} ) sgn(w,w',w'') \\ \nn
     & = & \frac{1}{16 \times 3!} \sum_{w,w',w'' \supset v} \epsilon_{IJK} e_w^I e_{w'}^J e_{w''}^K sgn(w,w',w'').
\eeqa
\\
This is the discretization of the integrated volume three-form $\int_{v} Tr ({\bf e} \wedge {\bf e} \wedge {\bf e})$ appearing in the mass term of the Dirac action. Once again, $sgn(w,w',w'')=\pm1$ here depending on the sign of the volume generated by the three vectors $e_w, e_{w'}, e_{w''} \in \mathbb{R}^3$: it carries the sign of the orientation of the vectorial basis formed by the three vectors. The numerical factor comes from the fact that the volume of the tetrahedron dual to the vertex $v$ appears more than once in the above expression \cite{GF}. Indeed, there are $C_6^3=20$  different triples of vectors constructable from the six faces dual to the six edges of a given tetrahedron \footnote{$C_n^p=\left( \begin{array}{c} n \\ p \end{array} \right)$ is the binomial coefficient.}. Out of these twenty triples, only sixteen, without counting the permutations, do not share a dual edge, i.e span the volume of the tetrahedron dual to the vertex $v$. We call these triples admissible. For each such triple, there are $3!=6$ possible permutations and hence $16 \times 3!$ times the volume of the three-simplex dual to the vertex $v$ in expression \eqref{volume}.

The presence of this volume term in the simplicial theory has dictated the discretization procedure that we have followed: we have suppressed the centers $c \in \mathcal{V}_+$ of the faces and the vertices of type $v_+ \in \veb$ as a possible habitat for the fermions. The reason is that the centers are not topologically dual to a three-dimensional region and the vertices of type $v_+ \in \veb$ belong to a given triple of wedges which is not admissible, i.e it does not generate a three-volume.  
We accordingly see how the rigidity of the structure of the Dirac action forces us to consider a particular type of discretization procedure. The fermions {\itshape can only be assigned} to the vertices $\ve$ of the dual complex $\T^*$
and by association the spin connection can only map the edges $\e$ of $\T^*$ to the spin group $\g$.   

The fermionic part \eqref{Daction} of the simplicial action is a polynomial function (in the sense of exterior multiplication) from the simplicial field configurations into the second exterior power $\Lambda^2 ( E \oplus \overline{E}^* )$ of the Grassmann algebra $\G$. We can exhibit the quadratic structure of the fermionic action via a compact expression by introducing a generalised Hermitian form, i.e a $\G$-valued, $\G$-module morphism 
$( , ):\mathbb{S} \times \mathbb{S} \rightarrow \G$ for which the chosen basis of $\mathbb{S}$ is orthonormal:  
\beq
S_{D,\T}[{\bf e}_w, g_e, \overline{\psi}_v, \psi_v]= (\psi, W({\bf e}_w, g_e) \psi) =
\sum_{uv} \overline{\psi}_u 
W_{u v}({\bf e}_w, g_e) \psi_v \h \in \h \bigwedge \hspace{.00000001mm} ^2 (E \oplus \overline{E}^*), 
\eeq
\\
where 

\beq
\label{Wmatrix}
W^{\hspace{2.5mm} \alpha}_{u v \hspace{1mm} \beta} = \frac{3i}{8} \frac{1}{m} ( A_{I \hspace{.5mm} uv} (\sigma^I U_{uv})^{\alpha}_{\hspace{1mm} \beta} -  
A_{I \hspace{.5mm} vu}  (U^{\dagger}_{vu} \sigma^I)^{\alpha}_{\hspace{1mm} \beta}) 
- \frac{1}{2} V_u \hspace{.5mm} \delta_{u v} \hspace{.5mm} \delta^{\alpha}_{\hspace{.5mm} \beta} \hspace{2mm} \in \C \subset \mathcal{G}_{\T}.
\eeq
\\ 
are the complex coefficients in the decomposition of the element $S_{D,\T}$ in the basis $\{\overline{\psi}_{u \alpha} \wedge \psi_v^{\beta}\}_{uv,\alpha\beta}$ of $\Lambda^2 (E \oplus \overline{E})$ given by the matrix elements of the $\mathcal{G}_{\T}$-module homomorphism $W : \mathbb{S} \rightarrow \mathbb{S}$ on the space of simplicial spinors. Note that we have rescaled the fermionic fields $\psi \rightarrow \sqrt{m/3} \psi$ such that $[\psi]=m^{3/2}$.

To prove that the simplicial action converges towards the correct continuum limit we procede in two steps. First, by injecting the first order expansions of the discretized fields in the discrete gravitational action \eqref{GRaction} and by using the fact that the simplicial curvature associated to a given wedge $w$ approximates the curvature at first order in the area of $w$, it is straight forward to see that $S_{GR,\T} \rightarrow - S_{GR}$ pointwise when $n \rightarrow \infty$ and the size of each tetrahedron shrinks to zero uniformly in the formal symbol $k$ labelling the three-simplices.

Secondly, we analyse the fermionic sector $S_{D,\T}$. Once again, we expand the simplicial fields in small vertice to vertice inter-distances corresponding to a high degree of refinement $(n >> 1)$ of the triangulation $\T$. The Taylor expansion of the fermionic fields around the vertex $u \in \ve$ yields, in the coordinates associated to the patch $U \subset M$ containing the vertex $u$ together with its four boundary tetrahedra,

\beq
\psi_v = \psi_u + \Delta_{uv}^{\mu} \partial_{\mu} \psi_u + \mathcal{O}(\Delta x^2).
\eeq
\\
By using Stokes theorem, it is furthermore possible to show that for a fixed vertex $u \in \ve$, the sum of the four coordinate area bivectors corresponding to the four triangles dual to the four edges emerging from $u$ sum to zero \cite{lattice}. This is due to the fact that the integral of the unit normal pointing outwards to a two-sphere piecewise linearly embedded in $\mathbb{R}^3$ must vanish. Explicitly,

\beq 
\label{normals}
\forall u \in \ve, \hspace{3mm} \sum_v \left( v_{ij} \wedge v_{ik} \right)^{\mu \nu}_{ijk=uv^*} =0,
\eeq
\\
where $ijk$ is the positively oriented triangle dual to the half-edge $uv \in \e_-$.

Finally, we use the fact that the normal coordinate area one-form to a given triangle $\Delta^2$ of $\mathcal{T}$ and the dual length vector going through $\Delta^2$ generate a coordinate three-volume \cite{discreteaction1} :

\beq
\label{coord}
\sum_{uv} A_{\mu \hspace{.5mm} uv} \Delta x_{uv}^{\nu} = 2 \Omega(M) \hspace{.5mm} \delta^{\nu}_{\mu} ,
\eeq 
\\
where $\Omega(M)$ is the coordinate volume of the spacetime manifold $M$ and the normal one form $A_{\mu \hspace{.5mm} uv}$ is related to the area bivector $(v_{ij} \wedge v_{ik})^{\mu \nu}$ of the triangle $ijk$ dual to the edge $uv$ through a (coordinate) hodge operation: $(v_{ij} \wedge v_{ik})^{\mu \nu}=\epsilon^{\mu \nu \rho} A_{\rho \hspace{.5mm} uv}$.

By inserting the Taylor expanded fields in the discretized action \eqref{Daction}, using expression \eqref{normals} to kill the undesired terms and by finally recovering the coordinate volume trough \eqref{coord}, we can conclude that, if the sequence of simplicial manifolds considered here converges to a smooth manifold with spin structure, the simplicial action \eqref{Daction} converges pointwise towards (minus) the Einstein/Dirac continuous action when the characteristic length of the simplices gos to zero and the number of simplices tends to infinity, at least formally. 

We finally verify that the action \eqref{Daction} is real as it is the case in the continuum.

\subsection{Fermionic integration}

We are now in position to give sense to the formula \eqref{pathintegral} by using the discretization prescription discussed above. The main obstacle to analytically work with path integrals resides in the definition of the functional integration measure. The lattice regularization used here provides us with an effective tool towards such a definition. 

\newpage

\subsubsection{Berezin integral and fermionic determinant}

The formal expression \eqref{pathintegral} becomes:

\beq
\label{partition}
\mathcal{Z}(\mathcal{T})=\left( \prod_{\W} \int_{\hat{\mathfrak{g}}} d {\bf e}_w \right) \left( \prod_{\e_-} \int_{\hat{G}} d u_{e_-} \right) \left( \prod_{\e_+} \int_{\hat{G}} d g_{e_+} \right)
\left( \int_{\G} d \mu (\overline{\psi}_v, \hspace{1mm} \psi_v) \right) \hspace{1mm} e^{iS_{\T}[{\bf e}_w,g_e,\overline{\psi}_v,\psi_v]},
\eeq 
\\
where $d {\bf e}_w$ is the Lebesgue measure on $\hat{\mathfrak{g}}=\mathfrak{spin}(3) \simeq \mathbb{R}^3$, 
$d g_e$ ($du_e$) is the normalised left Haar measure on $\hat{G}=Spin(3)$ (which coincides with the right Haar measure because $\hat{G}$ is compact: in this sense, $\hat{G}$ is said to be unimodular) and the symbol $d \mu (\overline{\psi}_v, \hspace{1mm} \psi_v) $ denotes the Berezin integral \cite{berezin} on $\mathcal{G}_{\T}$. The Berezin integral is the element of the algebraic dual $\mathcal{G}_{\T}^*$ of the Grassmann algebra $\mathcal{G}_{\T}$ defined by 

\beq
\forall a \in \mathcal{G}_{\T}, \hspace{2mm} \int_{\G} a \hspace{1mm} d \mu (\overline{\psi}_v, \hspace{1mm} \psi_v) := \left( \prod_{v=n}^{1} \prod_{\alpha=2}^{1} \frac{\partial}{\partial \overline{\psi}_{v\alpha}} \right)\left( \prod_{v=n}^{1} \prod_{\alpha=2}^{1} \frac{\partial}{\partial \psi_v^{\alpha}} \right) \hspace{1mm} a \hspace{2mm} \in \C.
\eeq
\\
As a direct consequence of the above formula, the Berezin integral of the exponential of a quadratic expression in the Grassmann variables yields the (totally antisymmetric) coefficient in front of the basis element $\prod_{v,\alpha} \psi_v^{\alpha} \prod_{v,\alpha} \overline{\psi}_{v \alpha}$ in the (finite) development of the exponential. This means that  
the Berezin integral in \eqref{partition} produces a sum over all possible alternating $2n$-th powers of the coefficient $W$, i.e the determinant of 
$W$ regarded as an endomorphism of the $\mathcal{G}_{\T}$-module $\mathbb{S}$. 
This exact manipulation is due to the fact that we have cut down the number of degrees of freedom of the theory from infinite to finite by introducing a lattice discretization.
The computation of the Berezin integral appearing in the partition function follows

\beq
\label{determinant}
\int_{\G} d \mu (\overline{\psi}_v, \hspace{1mm} \psi_v) e^{i (\overline{\psi}, W \psi)} = det \hspace{1mm} W,
\eeq
\\
and we can accordingly integrate out the fermionic degrees of freedom of the theory hence obtaining the usual path integral for gravity with insertion of an observable corresponding to the functional determinant $det \hspace{1mm} W$:

\beqa
\mathcal{Z}(\mathcal{T}) &=& < det \hspace{1mm} W >_{\mathcal{T}} \hspace{1mm} \\ \nn
&=& \left( \prod_{\W} \int_{\hat{\mathfrak{g}}} d {\bf e}_w \right) \left( \prod_{\e_-} \int_{\hat{G}} d u_{e_-} \right) \left( \prod_{\e_+} \int_{\hat{G}} d g_{e_+} \right) \left[det \hspace{1mm} W ({\bf e}_w, g_e) \right] e^{i \sum_w Tr ({\bf e}_w U_w)}.
\eeqa
\\
We now develop a tool to calculate the functional determinant $det \hspace{1mm} W$ leading to the computation of the path integral of the model.

\subsubsection{Fermionic loop expansion}

In statistical physics, the high-temperature expansion is a valuable tool for investigating critical phenomena. In lattice-like theories incorporating massive fermions, an analogous procedure called the hopping parameter expansion \cite{hopping} exists. The technique consists in expanding in inverse fermionic masses in the heavy mass regime of the theory. Of course, the sole consideration of the lowest orders in the expansion is reliable only if the fermions are heavier than the energy scale of the other fields of the theory under study. In the case of quantum gravity, the typical energy scale is given by the Planck mass which implies that the expansion parameter turns out to be large. However, if the expansion is finite and considered as a computational tool and not as a perturbation theory, we believe that the hopping parameter expansion in quantum gravity remains a valuable tool. This idea has in fact successfully been implemented in \cite{hoppingGR} where the coupling of massive fermions to two dimensional Lorentzian dynamical triangulations is considered. The large value of the expansion parameter simply implies that only a complete consideration of all orders will yield a meaningfull result. In other words, the hopping parameter expansion can only be used in quantum gravity as a computational technique, not as an effective description of the theory: we will not construct a perturbation theory but use the finite hopping parameter expansion to compute explicitly the partition function of the theory.  
The crystal clear geometrical interpretation of the hopping parameter expansion in terms of fermionic loops where the fermionic couplings to gravity appear as extremely natural and the explicit solvability of the model comfort us in the idea that this technique is an appealing tool to couple matter to gravity in the quantum regime, at least on a fixed triangulation.
Indeed, it is important to stress that the finiteness of the expansion that we are about to derive is due to the regularization defined by the fixed triangulation. The behaviour of the expansion under refinement of the triangulation is a question that we will leave open. Leaving this issue aside, we now show how to compute the partition function of the theory order by order in the expansion parameter obtaining a reliable evaluation of the fermionic determinant.

The functional determinant that we wish to calculate is defined by equation \eqref{determinant}. The first step is to realise a crucial symmetry of the $W$ endomorphism under charge conjugation. The charge conjugate spinors 

\beq
\psi_c = C^{-1} \psi^* \hspace{3mm} \mbox{and} \hspace{3mm} \overline{\psi}_c = \overline{\psi}^* C ,
\eeq
\\
where $C$ is the quaternionic structure defined in the appendix and $^*$ denotes complex conjugation, are solutions to the Dirac equation with charge sign inversion $e \rightarrow -e$ (when coupled to an electromagnetic field). 

The Clifford representation $\rho:\mathcal{C}(3) \rightarrow End (\mathbb{C}^2)$ is equivalent to the Hermitian conjugate representation $\rho^{\dagger}$ acting on $\bar{\C}^{2*}$ via the bijective intertwining operator \footnote{An intertwining operator $\phi$ between two representations $\pi_V$ and $\pi_W$ of $G$ acting respectively on the vector spaces $V$ and $W$ is a linear map between $V$ and $W$ satisfying $\phi \circ \pi_V = \pi_W \circ \phi$. The space of intertwiners is a vector space noted $Hom_G(V,W)$.}

\beq
A = \overline{\epsilon} \circ C : \C^2 \rightarrow \bar{\C}^{2*},
\eeq
\\
where $\epsilon :  \C^2 \rightarrow \C^{2*}$ is the two dimensional symplectic form ($\bar{\epsilon} : \bar{\C}^2 \rightarrow \bar{\C}^{2*}$ is the complex conjugate map) whose matrix representation is given by the totally antisymmetric tensor 
\beq
\epsilon = \left( \begin{array}{cc} 0 & 1 \\
                                  -1 & 0 \end{array} \right).
\eeq
\\ 
This vector space isomorphism mapping $\C^2$ onto its contragredient representation space realizes the equivalence between the two representations. Since $A$ is the identity map, $\rho(\gamma)^{\dagger} = \sigma^{\dagger} = \sigma$, we find that the charge conjugation matrix $C$ is given by the inverse of the dual pairing map $\epsilon$. 

Putting indices, the conjugate spinors and co-spinors are defined by

\beq 
\psi_c^{\alpha} = \epsilon^{\alpha \beta} \h \overline{\psi}_{\beta} \hspace{3mm} \mbox{and} \hspace{3mm} \overline{\psi}_{c \alpha} = - \psi^{\beta} \epsilon_{\beta \alpha},
\eeq
\\
where $\epsilon_{12}=\epsilon^{12}=1$ and $\epsilon_{\alpha \beta} \epsilon^{\beta \gamma} = -\delta_{\alpha}^{\gamma}$. One can verify that three dimensional Euclidean charge conjugation is indeed an anti-involution, preventing us from defining true Majorana spinors. 

These results trivially generalise to the case at hand namely the simplicial, spacetime dependent, Grassmann valued case.
However, $C$ now becomes a $\G$-module morphism $C: \mathbb{S} \rightarrow \overline{\mathbb{S}}$, where the conjugate module $\overline{\mathbb{S}}$ is defined with an appropriate real structure on the complex vector space $\G$.

Remarkably, the fermionic endomorphism $W$ transforms as

\beq
\label{symmetry}
C W_{uv}^{t} C^{-1} = W_{vu} \hspace{2mm} \Leftrightarrow \h \epsilon W_{uv} \epsilon = - W_{vu}^t ,
\eeq
\\
under charge conjugation.

In order to exploit this symmetry, we perform an affine transformation on $\C^2$ and introduce a family $\phi_{i}$, $i=1,2$, of symplectic Majorana spinors (resp co-spinors) :

\beq
\label{majorana}
\phi_{ic} = \epsilon_{ij} \phi_j, 
\eeq
\\
defined by

\beq
\begin{array}{cccccc} \phi_1 & = & \frac{1}{2} ( \psi_c + i \psi ) , \hspace{5mm} & \overline{\phi}_1 & = & \frac{1}{2} ( \overline{\psi}_c - i \overline{\psi} ) \vspace{2mm} \\
\phi_2 & = & -\frac{i}{2} ( \psi_c - i \psi ) , \hspace{5mm} & \overline{\phi}_2 & = & \frac{i}{2} ( \overline{\psi}_c + i \overline{\psi} ).
             \end{array}  
\eeq
\\
The components of these new fields are not independent.
The symplectic Majorana condition \eqref{majorana} implies the following set of constraints on the field components

\beq
\begin{array}{cc} \phi_1^{\alpha} = & - \epsilon^{\alpha \beta} \h \overline{\phi}_{2\beta} \\
\phi_2^{\alpha} = & \epsilon^{\alpha \beta} \h \overline{\phi}_{1\beta}
             \end{array} 
\eeq
\\
Hence, we have not modified the number of degrees of freedom involved in the theory. As we are about to see, the introduction of the symplectic Majorana fields however simplifies the situation rather drastically.
 
Indeed, reexpressing the Dirac action in terms of the new fields, using the conjugation symmetry \eqref{symmetry}, the above constraints and the Grassmann nature of the field components, we see that the two families decouple 

\beq
S_{D,\T}[\overline{\psi}_v, \psi_v] =
\sum_{u,v} \overline{\psi}_u W_{u v} \psi_v = -i (S_{\T}[\phi_{1v}] + S_{\T}[\phi_{2v}]),
\eeq
\\
where 
\beq
S_{\T}[\phi_v]:=S_{\T}[{\bf e}_f,g_e,\phi_v]=\sum_{uv} \phi_u^{\alpha} W_{uv \alpha \beta}({\bf e}_w, g_e) \phi_v^{\beta} ,
\eeq
\\
and $W_{uv \alpha \beta} = \epsilon_{\alpha \gamma} W^{\hspace{2.5mm} \gamma}_{u v \hspace{1mm} \beta}$. This action is very similar to the original simplicial Dirac action except that the degrees of freedom are reduced by a factor two via the symplectic Majorana constraints. The sum of the two actions however restores the original degrees of freedom, they have simply decoupled. This procedure gives us the opportunity of calculating the fermionic determinant exactly. With the original framework in terms of Dirac spinors, the presence of ``too many'' degrees of freedom per spacetime point prevents us from computing the terms at all order in the expansion that we now define.
 
By writing the integral definition of the fermionic determinant \eqref{determinant} in terms of the new fields, we obtain

\beq
det \h W = Pf(\epsilon W)^2 ,
\eeq
\\
where
\beq
Pf(\epsilon W)= \frac{1}{2^n} \int_{\Lambda(E)} d \mu (\phi_v) e^{S_{\T}[{\bf e}_w, g_e,\phi_v]},
\eeq
\\
is the Pfaffian \footnote{The Pfaffian of an antisymmetric matrix $M_{ab}$ of rank $2n$ is defined as follows. Let $M=\sum_{a<b} M_{ab}\;e_a\wedge e_b$,  
where $\{e_a\}_{a=1,...,2n}$ denotes a basis of $\R^{2n}$, be the associated bivector. The Pfaffian $Pf(M)$ is then defined by the equation :
$\frac{1}{n!}M^{\wedge n} = Pf(M) \;e_1\wedge e_2\wedge\cdots\wedge e_{2n}.$} 
of the antisymmetric endomorphism $\epsilon W$, $W_{uv \alpha \beta}=-W_{vu \beta \alpha}$, whose square yields the determinant of $\epsilon W$ equating $det \h W$ because the determinant is a morphism of endomorphism algebras.
Note that the Grassmann algebra integrated over is now ``half'' the original Grassmann algebra $\G$, namely the exterior algebra over $E$. Furthermore, we have chosen as a basis for $E$ the components of the symplectic Majorana field $\phi$.

We now first concentrate on the calculation of the Pfaffian of the endomorphism $\epsilon W$ before computing the square. The idea is to construct a finite perturbative definition of the integral in terms of inverse fermionic masses. We rewrite the Pfaffian by splitting the exponent into the volume contribution and the kinetic term of the $W$ endomorphism (see \eqref{Wmatrix}). The last is then expanded in inverse fermionic masses. We obtain the finite expansion

\beq
\label{expansion}
Pf(\epsilon W)= \frac{1}{2^n} \sum_{k=0}^n \alpha^k \Gamma_k,
\eeq
\\
where the coefficients $\Gamma_k:=\Gamma_k({\bf e}_w,g_e)$ are given by 
\beq
\label{orderk}
\Gamma_k = \frac{1}{k!} \sum_{u_1v_1,...,u_kv_k} \int_{\Lambda(E)} d \mu (\phi_v) \prod_v e^{-\frac{1}{2} V_v \phi_v \epsilon \phi_v} 
\left[ \phi_{u_1} \epsilon D_{u_1v_1} \phi_{v_1} ... \phi_{u_k} \epsilon D_{u_kv_k} \phi_{v_k} \right],
\eeq
\\
with the kinetic term $D$ :
\beq
\label{Dmatrix}
D_{uv}=\frac{3i}{8}\left( A_{Iuv} \sigma^I U_{uv} - A_{Ivu} U^{\dagger}_{vu} \sigma^I \right).
\eeq
\\
Putting back physical units ($c$ is left fixed to unity), the parameter $\alpha$ is given by $\alpha=\frac{\hbar}{m l_p}=\frac{m_p}{m}$ with $l_p=G \hbar$ and $m_p=\frac{1}{G}$ respectively denoting the Planck length and mass. To explicitly realise the appearance of the Planck length, we have rescalled the triad such that $e \rightarrow \frac{1}{G \hbar} e$. Finally, the passage from the exponential of a sum to the product of exponentials is due to the sole presence of even elements of $\Lambda \left( E \right)$ in the exponent.
  
We are now ready to perform the integrals order by order in $\alpha$. Most terms in the expansion vanish after integration. The Berezin integral yields a non-vanishing result if and only if each vertex $v \in \ve$ is exactly saturated by two (non identical) symplectic Majorana fermionic components. As a result, each vertex either carries a volume contribution or must be a source {\itshape and} a target point for a fermionic worldline. 
For instance, we obtain the following identity

\beq
\label{formula}
\int_{\Lambda(E)} d \mu (\phi_v) \prod_v e^{-\frac{1}{2} V_v \phi_v \epsilon \phi_v} \phi_t^{\alpha} \phi_u^{\beta} = (-1)^{n-1} \prod_{v\neq u} V_v \delta_{tu} \epsilon^{\alpha \beta}.
\eeq
\\
The generalisation of the above formula to an arbitrary number $2k$ of field insertions implies that, leaving aside the sum over vertices, the resulting term in the Berezin integral of the coefficient of order $k$ \eqref{orderk} is a contribution consisting in a closed sequence of products of kinetic terms $D$ going through $k$ vertices of the dual complex $\T^*$ together with volume weights living everywhere in the complementary of the sequence. These loops can close on a vertex $u$ and reopen on the neighbooring vertex $v$ yielding disconnected sequences such that the number of vertices traversed by the connected components sum up to $k$. A sum over the $k$ vertices supporting the closed product is then implemented by the sum over vertices in equation \eqref{orderk}.  

Forall $k \in ]0,n[$ and for any fixed set $\ve_k=\{v_1,v_2,...,v_k\} \subset \ve$ of $k$ two by two adjacent, distinct vertices, such a contribution, noted $\Gamma_k^{(v_1,...,v_k)}$, will be defined in the connected case by

\beq
\Gamma_k^{(v_1,...,v_k)} = (-1)^{k-1} 
(-1)^{n-k} \prod_{v \notin \ve_k} V_v \h Tr_D \left( D_{v_1 v_2} D_{v_2 v_3} ... D_{v_k v_1} \right),
\eeq
\\
where $Tr_D: End(\mathcal{S}(\ve) \otimes \V) \rightarrow End(\mathcal{S}(\ve))$ is the partial trace acting on the Dirac part of $End(\mathbb{S})$. If the sequence $\Gamma_k^{(v_1,...,v_k)}$ is not connected, there will be one Dirac trace appearing in the above formula for each connected component. 

The coefficient $\Gamma_k$ is accordingly a sum over connected and/or disconnected contributions :

\beq
\label{sumloops}
\Gamma_k = \frac{1}{k!} \sum_{\ve_k}\sum_{\Gamma} \Gamma_k^{(v_1,...,v_k)}, 
\eeq
\\
where the first sum is over possible vertices supporting the loop and the second formal sum symbol
\footnote{More precisely, by denoting $\tilde{\gamma}_{\ve_k}$ the Dirac trace $Tr_D(D_{v_1 v_2} D_{v_2 v_3} ... D_{v_k v_1})$,
we could define the coefficient $\Gamma_k$ by introducing a partition $\Pi_p$ of $\ve_k$ with $p$ subsets: $\ve_k=\cup_{i=1}^p \ve_{k_i}$, $\sum_{i=1}^p k_i=k$, in which case 

$\Gamma_k = \frac{1}{k!} \sum_{\ve_{k}} \prod_{v \notin \ve_k} \sum_{p=1}^{k} \sum_{\Pi_p} \bigcup_{i=1}^p \tilde{\gamma}_{\ve_{k_i}}$,

where the first sum is over all vertices, the second over all numbers of subsets $p$ constituting the partition and the last sum is over all possible partitions with $p$ subsets.}
implies a sum over all possible connected and/or disconnected contributions saturating the $k$ vertices $\ve_k$.   

We now compute the extremal configurations. The order zero is the no loop case and is given by  

\beq
\Gamma_0 = (-1)^n \prod_{\ve} V_v.
\eeq
\\
Note that the first order yields the trace of the $D$ endomorphism by virtue of equation \eqref{formula} which is null. 

If the order of the loop $k$ equates the number of vertices $n$, there is no place for the volume contributions and the order $k$ coefficient $\Gamma_k$ in the expansion \eqref{expansion} yields 

\beq
\Gamma_n = 2^n (-1)^{n-1} Tr_D \left( D_{v_1 v_2} D_{v_2 v_3} ... D_{v_n v_1} \right).
\eeq 
\\
Once again, the loop can be disconnected and built out of lower order connected components in which case there is one Dirac trace per connected piece.
What do these loops have to do with vacuum fermionic loops ?
Each Dirac trace over a closed sequence of $D$ endomorphisms generates a sum over $2^k=\sum_{p=0}^k C_k^p$ different contributions by virtue of equation \eqref{Dmatrix}. For a fixed set $v_1,v_2,...,v_k$ of $k$ vertices,

\beq
\label{sumconfigurations}
Tr_D \left( D_{v_1v_2} D_{v_2v_3} ... D_{v_kv_1} \right) = (\frac{3i}{8})^k \sum_{i=1}^{2^k} \gamma^{i \hspace{.5mm} (v_1 v_2 ... v_k)}_k.
\eeq  
\\
Each element $\gamma$ in the above sum is a closed fermionic path going trough the $k$ vertices $v_1,...,v_k$. Such a loop is attached to a decoration, i.e a pattern of sigma matrices insertions together with a polynomial function of the discretized triad $e_w$.

\begin{figure}[t]
\begin{center}
\includegraphics{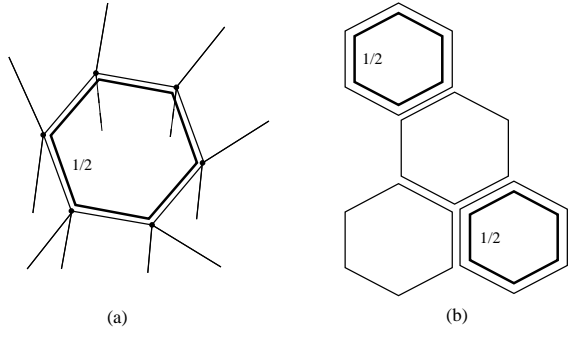}
\caption{(a) A fermionic loop of order six. (b) An order twelve product of two loops.}
\end{center}
\end{figure}

Let us introduce some terminology. As suggested by the notation above,
we note $\gamma_k$ a loop wrapping around $k$ vertices of $\T^*$. Such a loop will be called of order $k$. All fermionic loops are connected, do not go trough the same vertex more than once and when a product of loops is generated at given order in the expansion by a contribution $\Gamma_k^{(v_1,...,v_k)}$, the different loops do not touch each other (see fig.1). 
For each fixed set of $k$ vertices $v_1,...,v_k \in \ve$, the contributions appearing in the above sum \eqref{sumconfigurations} differ between themselves by their decorations, i.e by the positioning of the sigma matrices inside the trace and their pattern of contraction with the (internal) area co-vectors $A_I$. We call such a contribution $\gamma^i$ in the above sum a decorated loop and note $\mathcal{L}_k$ the set of decorated $k$-loops. We define an equivalence relation $\sim$ on the set $\mathcal{L}_k$ of loop configurations of order $k$ by calling two contributions equivalent if they can be mapped onto each other by relabelling of the vertices. This is the case when the two decorated loops traversing the same set of vertices are built out of the same sequence of patterns alternating the sigma matrices and the group elements inside the Dirac trace up to cyclicity.

A typical generic decorated loop of order $k$ looks like this: $\forall v_i \neq v_j$,

\beq
\label{configuration}
\gamma^{(v_1 v_2 ... v_k)}_k = A_{I_1 \hspace{.5mm} v_1v_2} A_{I_2 \hspace{.5mm} v_2v_3} ... A_{I_k \hspace{.5mm} v_kv_1} \h Tr_D \left( \sigma^{I_1} U_{v_1v_2} \sigma^{I_2} U_{v_2v_3} ... \sigma^{I_k} U_{v_kv_1} \right).
\eeq
\\
This term is the first in the sum over decorations \eqref{sumconfigurations} and will be the generic loop with which we will work to build the model. All of the other decorated loops of a given order $k$ can be obtained and usefully classified trough the following prescription. 

The traces \footnote{We are leaving the area co-vectors out of the discussion for the sake of simplicity.} are built out of a succession of two types of patterns of the form $\sigma U$ and $U^{\dagger} \sigma$ (see \eqref{Dmatrix}). We call the first pattern (resp second pattern) a pattern of type $a$ (resp of type $b$). A contribution of order $k$ will contain a succession of $k-m$ (resp $m$) type $a$ (resp type $b$) patterns. There are $C_k^m$ possible combinations of such type which can be classified as follows. First, represent each pattern by a labelled point on a circle (type $a$ or type $b$) and each combination of such patterns (up to cyclic permutation) by a set of labelled points on a circle. Changing the starting point on the circle corresponds to a relabelling of the vertices. Hence, each circle corresponds to an element in $\mathcal{L}_k/\sim$. Our classification is established by identifying the finite number $N$ of possible combined patterns of the form ``type $b$ times type $a$'', i.e  $U^{\dagger} \sigma \sigma U$, constructable out of the fixed number $k-m$ (resp $m$) of type $a$ (resp type $b$) patterns. For each fixed $N$, there are $N$ sequences of the form $\sigma U \sigma U ... \sigma U U^{\dagger} \sigma U^{\dagger} \sigma ... U^{\dagger} \sigma$ constituting each circle. The different possible sequences for fixed $N$ correspond to the different $\sim$-inequivalent contributions containing $N$ patterns of two adjacent Pauli matrices inside the Dirac trace. We will call the sub-sequences of the form $\sigma U \sigma U ... \sigma U$ 
(resp $U^{\dagger} \sigma U^{\dagger} \sigma ... U^{\dagger} \sigma$) type $a$ (resp type $b$) sub-sequences. 
   
The contributions to the Pfaffian \eqref{expansion} have now been defined and classified.
To summarise, the calculation of the Pfaffian yields a sum over products of connected decorated loops wrapping around more and more vertices of the dual triangulation $\T^*$ with the increasing of the order of the expansion. This sum stops with an extremal decorated loop saturating all the vertices of $\T^*$. At given order $k$, the coefficients in the expansion are symbolically written

\beq
\Gamma_k = \sum_{\bar{\gamma}_k} c_{\bar{\gamma}_k} \prod_{v \notin \bar{\gamma}_k} V_v \h \bar{\gamma}_k, \hspace{5mm} c_{\bar{\gamma}_k} \in \C ,
\eeq
\\
where the sum is over all possible collections $\bar{\gamma}_k := \cup_i \gamma_i$ of decorated loops $\gamma$ contributing to the order $k$. Our conventions are such that $\bar{\gamma}_0 = \gamma_0 = 1$. We call the ensemble $\bar{\gamma}_k$ a loop configuration.
The complex number $c_{\bar{\gamma}_k}$ contains all the numerical factors including the multiplicity of the loop.

We can now compute the square of the Pfaffian to obtain the functional determinant $det \h W$ and restore all the degrees of freedom of the (discretized) theory. The determinant expansion will accordingly involve a sum over quadratic terms in the loop configurations. At a given order $k$, leaving aside the sum over vertices, we obtain terms of the form $\Gamma_{p+q}=\Gamma_p . \Gamma_q$ with $p+q=k$. We distinguish between three different types of contributions. 

First, there are the terms where both $p$ and $q$ are smaller that the number $n$ of vertices in which case the product of loop configurations generates configurations where all vertices in the complementary of the union of the two fermionic loop configurations are associated to square volumes while the vertices traversed by the fermionic loops carry a single volume contribution.
We can now find vertices where more than one fermionic line converges and edges supporting two fermionic lines. These particular vertices will be called degenerate and do not carry a volume weight. 
If $p$ is smaller that $n$ and $q$ equates $n$, the intersection between the two fermionic loop configurations is necessarily non empty and the resulting loop configuration is degenerate. All non degenerate vertices contribute a volume term while the degenerate ones will carry a trivial weight.  
Finally, if both $p$ and $q$ are equal to $n$, we obtain a contribution consisting in no volume term and a double fermionic loop going around all of the two complex.  

The structure of the expansion being established, we can rewrite the partition function of the simplicial theory as the following {\itshape finite} sum

\beq
\label{partition1}
\mathcal{Z}(\mathcal{T})= \frac{1}{2^{2n}} \sum_{k=0}^{2n} \alpha^k \hspace{.5mm} \mathcal{Z}(\Gamma_k, \T), 
\eeq
\\
Here, $\Gamma_k = \sum_{p,q} \Gamma_p \Gamma_q$, such that $p+q=k$, and $\forall k \in [0,2n]$,

\beqa
\label{partition2}
\mathcal{Z}(\Gamma_k, \T) &=& \sum_{\bar{\gamma}_k} c_{\bar{\gamma}_k} \mathcal{Z}(\bar{\gamma}_k, \T) \\ \nn
&=& \sum_{\bar{\gamma}_k} c_{\bar{\gamma}_k} \hspace{.1mm} \left[ \left( \prod_{\W} \int_{\hat{\mathfrak{g}}} d {\bf e}_w \right) \left( \prod_{\e_-} \int_{\hat{G}} d u_{e_-} \right) \left( \prod_{\e_+} \int_{\hat{G}} d g_{e_+} \right) \times \right. \\ \nn
&& \left. \hspace{1mm} \left[ \prod_{\ve} f_{v,\bar{\gamma}_k} ({\bf e}_w) \hspace{.5mm} \bar{\gamma}_k ({\bf e}_w, g_e) \hspace{.5mm} \right] e^{i \sum_w Tr (e_w U_w)} \right].
\eeqa
\\
where $\bar{\gamma}_k = \bar{\gamma}_p \cup \bar{\gamma}_q$ with $p+q=k$ and $f_{v,\bar{\gamma}_k}$ is a polynomial function of the triad whose exact nature (identity, volume, squared volume) depends on the vertex $v$ and on the type of decorated loop product $\bar{\gamma}_k$ according to the discussion above.

\subsection{Fermions and quantum gravity}

We are  now ready to perform the integrations in the path integral and solve the model at the quantum mechanical level. It is rather straight forward to compute the integration over the discretized connection, i.e over the group elements attached to the edges. The difficulty resides in the calculation of the integral over the simplicial triad field. The idea that we will follow, due to Freidel and Krasnov \cite{GF}, is to use the fact that any polynomial function of the discretized triad appearing in the partition function can be obtained by source derivation on a given generating functional.

\subsubsection{Generating functional} 

We firstly study a modelling case where the loop configuration $\bar{\gamma}_k$ is of the form $\gamma_k \cup \gamma_0$ where $\gamma_k := \gamma^{(v_1,...,v_k)}_k, \h \forall k < n,$ is the single, connected decorated $k$-loop defined by \eqref{configuration}. This prototype situation retains all the important features of the model and will lead us to derive the full model containing all the other types of loop configurations at the end of this paper. Troughout this section, we will assume that $k>2$. 

Regarding the framework developed above, the associated weight in the partition function is given by

\beqa
\label{contribution}
\mathcal{Z}(\bar{\gamma}_k, \T) &=& \left( \prod_{\W} \int_{\hat{\mathfrak{g}}} d {\bf e}_w \right) \left( \prod_{\e_-} \int_{\hat{G}} d u_{e_-} \right) \left( \prod_{\e_+} \int_{\hat{G}} d g_{e_+} \right) \times \\ \nn 
&& \left[ \prod_{v \notin \gamma_k} V_v^2 ({\bf e}_w)  
\prod_{v \in \gamma_k} V_v ({\bf e}_w) \prod_{e \in \gamma_k} A_{I_e , e}({\bf e}_w) \hspace{1mm} Tr_D \hspace{1mm} (\prod_{e \in \gamma_k} \sigma^{I_e} U_e) \hspace{1mm} \right] e^{i \sum_w Tr ({\bf e}_w U_w)}.
\eeqa
\\
The technology of generating functionals gives us the possibility to compute such an expression needed for the construction of the model. 

The idea is to rewrite the above quantity of interest as 

\beq
\label{model}
\mathcal{Z}(\bar{\gamma}_k, \T) =  \left[ \prod_{v \notin \gamma_k} \hat{V}_v^2 \prod_{v \in \gamma_k} \hat{V}_v \prod_{e \in \gamma_k} \hat{A}_{I_e , e}  \left( \mathcal{Z}^{(I)}(J, \gamma_k, \T) \right) \right]_{J=0}.
\eeq
\\
Here, a degree $n$ element $\mathcal{O}$ of the algebra of monomial functions over $\mathfrak{spin}(3)^{\times n}$ is transformed into a source differential operator of order $n$ $\hat{\mathcal{O}}:=\hat{\mathcal{O}}(\frac{\delta}{\delta J})$ as follows

\beqa
\mathcal{O}({\bf e}_w) &\rightarrow& \hat{\mathcal{O}}(\frac{\delta}{\delta {J_w}}) \\ \nn
    e^I_w        &\rightarrow& \frac{i}{2} \h \frac{\delta}{\delta J_{I \hspace{.5mm} w}}.
\eeqa
\\
The generating functional $\mathcal{Z}^{(I)}(J, \gamma_k, \T)$ is a map from the space of $\mathfrak{spin}(3)$-valued two forms into $V^{\times k}$ defined by

\beqa
\label{source1}
\mathcal{Z}^{(I)}(J, \gamma_k, \T) &=& \left( \prod_{\W} \int_{\hat{\mathfrak{g}}} d {\bf e}_w \right) \left( \prod_{\e_-} \int_{\hat{G}} d u_{e_-} \right) \left( \prod_{\e_+} \int_{\hat{G}} d g_{e_+} \right) \times \\ \nn
&& \left[ Tr_D \hspace{1mm} (\prod_{e \in \gamma_k} \sigma^{I_e} U_e) \right] e^{i  S_{\T,J}[{\bf e}_w,g_e,e^{J_w}]},
\eeqa
\\
with $(I):=(I_{e_1},I_{e_2},...,I_{e_k})$ denoting a cumulative index and 

\beq
S_{\T,J}[{\bf e}_w,g_e,e^{J_w}] = \sum_w Tr \left( {\bf e}_w U_w e^{J_w} \right) ,
\eeq
\\
where $J_w:=\int_w J$, with $J = J^I X_I$, being a $\hat{\mathfrak{g}}$-valued current two-form. Using the traceless property of the elements of $\mathfrak{spin}(3)$, we see that $S_{\T,J}$ is a discretized version of the continuum source action 

\beq
S_J[{\bf e}, \hat{\omega}, J]=\int_M  Tr \h ( {\bf e} \wedge \hat{R} [\hat{\omega}] ) + Tr \h ( {\bf e} \wedge J ) ,
\eeq
\\
and that the operation

\beq
\frac{\delta}{\delta J_{Iw}} S_{\T,J}[{\bf e}_w,g_e,e^{J_w}] \mid_{J=0} \h = Tr ({\bf e}_w U_w X^I) \simeq -2 e_w^I ,
\eeq
\\
approximates the continuum operation of source derivation at first order in the segments $\Delta^1 \in \T$ lengths ($U_w \sim 1\!\!1$). This motivates our definition of the expectation value of any degree $n$ monomial in the simplicial triad :

\beq
< \mathcal{O}({\bf e}_w) >_{\T} \h := (\frac{i}{2})^n \left[ \left( \frac{\delta}{\delta J_{I w}} \right)^n \h \mathcal{Z}^{(I)}(J, \gamma_k, \T) \right]_{J=0} .
\eeq
\\
We now calculate the above generating functional \eqref{source1} to be able to make sense of equation \eqref{contribution} following the next lines \cite{GF}. 

\newpage

\paragraph{Simplicial triad integration.}

First, the Lebesgue integrations over ${\mathbb{R}^3} \simeq \gh$ can be performed yielding one delta function 
on the spin group 
\footnote{More precisely, these integrals produce delta functions on $G=SO(3)$ \cite{Freidel}. In fact, we are here considering integrals of the form

$$
\int_{\hat{\mathfrak{g}}} d {\bf e} \h e^{i Tr ({\bf e} g)}(1 + \epsilon(g)) = \delta(g),
$$
\\
where $\epsilon(g)$ is the sign of $\cos \theta$ in the parametrization $g(\theta,n) = \cos \theta \h 1\!\!1 + i \sin \theta \h n . \sigma$ of the spin group, as a mean to construct delta functions on $\hat{G} = Spin(3)$.
} 
for each wedge $w$.
Then, using the fact that the characters $\chi_j : \hat{G} \rightarrow \C$; $g \rightarrow Tr \stackrel{j}{\pi} (g)$, provide a vectorial basis for the space of central functions on $\hat{G}$ and that the delta function is central as a distribution, we obtain

\beqa
\label{source4}
\mathcal{Z}^{(I)}(J, \gamma_k, \T) &=& \left( \prod_{\e_-} \int_{\hat{G}} d u_{e_-} \right) \left( \prod_{\e_+} \int_{\hat{G}} d g_{e_+} \right) \times \\ \nn
&& \left[ Tr_D \hspace{1mm} (\prod_{e \in \gamma_k} \sigma^{I_e} U_e) \right] \prod_{\W} \sum_{j_w \in \mathbb{N}/2} dim(j_w) \chi_{j_w}(g_w e^{J_w}),
\eeqa
\\
where $dim(j_w)=2j_w + 1$ is the dimension of the representation $j_w$ assigned to the wedge $w$. 

\paragraph{Integration over the discretized connections.}

The next step consits in performing the Haar integrals. All $1$-cells belong to more than one $2$-cell. The $e_-$ type edges are shared by two wedges, the $e_+$ type belong to three or, as we shall see, four wedges or faces. The group elements assigned to the edges will therefore always appear in more than one trace in \eqref{source4}. We accordingly need to develop the necessary technique for integrating over $\hat{G}$ the tensor product of two, three or four representations of a given $g \in \hat{G}$.

Forall triple of unitary, irreducible representations $\stackrel{i}{\pi}, \stackrel{j}{\pi}, \stackrel{k}{\pi}$ of $\hat{G}$, let $\Psi^{k}_{\h ij}: \stackrel{i}{\V} \otimes \Vj \rightarrow \stackrel{k}{\V}$ and $\Phi^{ij}_{\h k} : \stackrel{k}{\V} \rightarrow \stackrel{i}{\V} \otimes \Vj$ denote the Clebsh-Gordan intertwining operators 
. These maps are uniquely defined up to normalisation since the vector space of three-valent $\hat{G}$-intertwiners is of dimension one.
The phase and the sign of the Clebsch-Gordan maps is fixed by requiring their reality, asking $(\Psi^{k}_{\h ij})^{\dagger}=\Phi_{k}^{\h ij}$ and by following Wigner's convention \cite{wigner}. We can define a Hermitian form $<,>$ on the vector space of intertwiners with respect to which the Clebsh-Gordans are normalised to unity $<\Psi,\Psi> = Tr (\Psi \Psi^{\dagger}) = Tr (\Psi \Phi)=1$.

We introduce the non canonical bijective intertwining operator 
\beqa
\label{dual}
\epsilon_j: \h \Vj   &\rightarrow& \overset{j}{\mathbb{V}}\,^* \\ \nn
\stackrel{j}{e}_a &\rightarrow& \epsilon_j (\stackrel{j}{e}_a) = \overset{j}{e}\,^b \epsilon_{j ba} ,
\eeqa
\\
with $\epsilon_{j ab}=(-1)^{j-b} \delta_{b,-a}$ and $\epsilon_{1/2}:=\epsilon$. This vector space isomorphism defines a scalar product on $\Vj$ via $(v,w)=(\epsilon(v))(w)$, $\h \forall v,w \in \Vj$, and proves that any representation $j$ is equivalent to its contragredient: $\overset{j}{\pi}\,^* = \h \stackrel{j}{\bar{\pi}} \h = \epsilon_j \stackrel{j}{\pi} \epsilon_j^{-1}$. Using this isomorphism, it is possible to raise and lower intertwiner indices i.e to construct, for instance, elements of $Hom_{\hat{G}}(\stackrel{i}{\V} \otimes \Vj, \stackrel{k}{\V^*})$ :
\beq
\label{dualintert}
\Psi_{kij} = \epsilon_k \circ \Psi^{k}_{\h ij}.
\eeq
\\

\newpage

We will extensively use a symmetrized version of the Clebsch-Gordan intertwiner called the Wigner three-$j$ map 

\beq
\iota_{ijk} : \h \stackrel{i}{\V} \otimes \stackrel{j}{\V} \otimes \stackrel{k}{\V} \rightarrow \C.
\eeq 
\\
Its relationship to the Clebsch-Gordan maps is given by the following evaluation
called a three-$j$ symbol.
Using the basis $\{\stackrel{l}{e_{a}}\}_a$ of $\stackrel{l}{\V}$, with $l=i,j,k$,

\beq
\iota_{ijk}(\stackrel{i}{e_{a}} \otimes \stackrel{j}{e_{b}} \otimes \stackrel{k}{e_{c}}) = 
\left( \begin{array}{lll}
i & j & k \\ 
a & b & c 
\end{array} \right) = e^{i \pi(i-j-k)} \h (\stackrel{k}{e_{c}} \h , \Psi^k_{\h ij} (\stackrel{i}{e_{a}} \otimes \stackrel{j}{e_{b}}) ) \h \in \R.
\eeq
\\
This intertwiner satisfies, among other properties, the symmetry conditions $\iota_{kij}=\iota_{ijk}=\iota_{jki}=e^{i \pi(i+j+k)} \iota_{ikj}$ and is normalised: $<\I,\I>=1$.

We are now ready to integrate out the group elements.    
The first step consists in taking care of the auxiliary variables $u$. This is achieved by remarking that this operation will require the integration over the tensor product of two representation matrices since the internal edges to which the auxiliary variables are attached are shared by precisely two wedges. 

Using the complete reducibility $\stackrel{i}{\pi} \otimes \stackrel{j}{\pi} \h = \Phi^{ij}_{\h k} \stackrel{k}{\pi} \Psi^{k}_{\h ij}$ 
of the tensor product of two irreducible representations, it is immediate to show that

\beq
\int_{\hat{G}} dg \stackrel{i}{\pi}(g) \h \otimes \stackrel{j}{\pi}(g) = 
\Phi^{ij}_{\hspace{1.5mm} 0} \Psi^0_{\h ij} ,
\eeq
\\
where $\Psi^0_{\h ij}(\stackrel{i}{e} \otimes \stackrel{j}{e})=\frac{1}{\sqrt{dim(j)}} (\epsilon_i (\stackrel{i}{e}))(\stackrel{j}{e}) = \frac{1}{\sqrt{dim(j)}} (\stackrel{i}{e} , \stackrel{j}{e})$ if $i$ and $j^*$ are equivalent and yields zero otherwise. 
The fact that the orientation of the wedges is induced by the orientation of the faces to which they belong implies that the two wedges meeting on a given edge $e_-$ will always have opposite orientations. Accordingly, the integration over the auxiliary variables will impose each wedge belonging to a given face $f$ to be coloured by the same representation $j_f$. Recalling that the representation associated to a wedge is interpreted as the quantum length number of the segment defining the center of the face containing the wedge as measured from the frame associated to the wedge, we obtain that the vector associated to the segment has the same quantum length in all frames \cite{GF}.

The generating functional \eqref{source1} consequently yields

\beqa
\label{source2}
\mathcal{Z}^{(I)}(J, \gamma_k, \T) &=& \prod_{\e} \int_{\hat{G}} d g_e \left[ Tr_D \hspace{1mm} (\prod_{e \in \gamma_k} \sigma^{I_e} U_e) \right] \times \\ \nn
&& \prod_{\f} \sum_{j_f \in \mathbb{N}} dim(j_f)^{\chi(f)} \chi_{j_f}( e^{J_f^1} g_{e_f^1} ... e^{J_f^m} g_{e_f^m}),
\eeqa
\\
where the group element insertions are along all the vertices $v \in \ve$ of a given face $f$ ($m$ denotes the number on vertices, i.e the number of wedges contained in $f$) and $\chi(f)$ denotes the topological invariant Euler characteristic of the face.  

The second step consists in Haar integrating over the group variables assigned to each edge $e$ of $\T^*$. The dimensionality of our spacetime implies that each edge $e$ is shared by exactly three geometrical faces (each triangle is built out of three segments). However, if a fermion is travelling along the edge $e$, there will be an extra face, that we will call virtual, in the fundamental $j=j_0=1/2$ representation corresponding to (a piece of) the fermionic loop.
We now distinguish between the two cases : the purely gravitational situation, away from the fermionic loop $\gamma_k$, and the edges supporting a fermion path. 

If no fermions are present, each group element will appear three times in the product over the faces in \eqref{source2}. If $i$, $j$ and $k$ denote the three geometrical colourings of the three positively oriented faces meeting on $e$, i.e the three lengths quantum numbers of the three segments building up the triangle dual to the edge $e$, the integral of the tensor product creates a three-$j$ symbols together with its adjoint intertwiner 

\beq
\label{intert}
\int_{\g} d g \stackrel{i}{\pi}(g) \otimes \stackrel{j}{\pi}(g) \otimes \stackrel{k}{\pi}(g)
=\iota^{\dagger} \h \iota ,
\eeq
\\
where $\iota := \iota_{ijk}$.
If one or two out of the three faces are not positively oriented, one uses the isomorphism \eqref{dual} on both sides of the above equation in order to obtain the integral of interest. The raising and lowering of indices on the right hand side yields the correct intertwiner via \eqref{dualintert}.

Let us now turn to the case where a fermion is travelling through the triangle dual to the edge $e$, i.e $e \in \gamma_k$.
There are now four representations associated to three geometric colourings of the three faces meeting on $e$ plus a matter colouring of the virtual face defined by the fermionic loop.

We are here contemplating the creation of quantum torsion by the fermionic field dislocating the quantum mechanical spacetime in which it is evolving. Indeed, if we consider the triangle $\Delta^2 \in \T$ traversed by the fermion line as a physical system determined by the (physical) lengths of its boundary segments, we can apply to it a Kirillov-Kostant geometric quantisation procedure obtaining the so-called quantum triangle \cite{qtriangle}. The lengths of the boundary segments of the quantum triangle are quantised and are encoded in the associated representations. When no spacetime torsion is present, the quantum version of the Gauss law tells us that the three quantum length vectors must close. In presence of fermions, the Gauss law acquires a source term. Heuristically, we see that the presence of the fourth virtual face is a gap in the closing of the quantum triangle whose magnitude is given by the associated representation, i.e $j_0=1/2$ in Planck units. The inspection of the classical equations of motion shows that this gap is the quantum analogue of the failure to the coplanarity of the image under the gravitational field $e$ in inertial space of the three (coordinate) length vectors of the triangle traversed by the fermion. 

In addition, equation \eqref{source1} teaches us that, for the generic case that we are considering here, each fermion path along an edge comes accompanied by a sigma matrix. Recall that, classically, the sigma matrices are matrix representations of the embedding of the co-inertial space $V^*$ in the Clifford algebra $\mathcal{C}(3)$ and hence solder between the spin space $\mathbb{V}$, i.e the $j=1/2$ representation space of $\g$, and the inertial space $V$, i.e the $j=1$ representation, created by the gravitational field at each point of spacetime (see the appendix).  
In the quantum framework developed here, we see that these representations of $\mathcal{C}(3)$ are attached to the edges of the fermionic loops. How do they relate the spin space to spacetime in the quantum regime ? The answer to this question resides in the rethinking of the sigma matrices as elements of $Hom_{\g}(V^*, \V \otimes \V^*)$ by virtue of the defining equation \eqref{spin} of the spin group $\g$.
Because $V^*$, $V$ and $\stackrel{1}{\V}$ are isomorphic as vector spaces, we can conclude that the sigma matrices are intertwining operators between the adjoint representation and the tensor product of a fundamental and its dual representation.

More precisely, by using the isomorphism $Hom_{\g} (V^* \otimes \V^*, \V^*) \simeq Hom_{\g}(V^*, End (\V^*))$ given by $\Phi (e^I \otimes e^{\alpha}) = \frac{1}{\sqrt{6}} (\sigma (e^I))^t (e^{\alpha})$, we obtain the following relation between sigma matrices and three-$j$ symbols 

\beq
\iota^{\dagger}(e^I \otimes e^{\alpha} \otimes e^{\beta}) = \left( \begin{array}{lll}
I & \alpha & \beta \\
1 & 1/2 & 1/2  
\end{array} \right) = (e^{\beta}, \Phi (e^I \otimes e^{\alpha})) = \frac{1}{\sqrt{6}} \sigma^{I \alpha}_{\hspace{2mm} \gamma} \epsilon^{\gamma \beta},
\eeq
\\
where we have omitted the representation labels of the intertwiners as there is no ambiguity and the scalar product used is the one induced on $\V^*$ par the previously defined scalar product on $\V$. This reformulation will represent the quantum analogue of equation \eqref{spin}.

Hence, when a fermion line $j_0=1/2$ comes attached to a sigma matrix, we will have to repeatidly use the following relation

\beq
\label{intert1}
\int_{\g} dg \h \sigma^I \h \circ \stackrel{j_0}{\pi}(g) \otimes \stackrel{i}{\pi}(g) \otimes \stackrel{j}{\pi}(g) \otimes \stackrel{k}{\pi}(g) 
= \sum_s \left( \sigma^I \circ \I^{\dagger}_s \right) \I_{s} ,
\eeq
\\
where $\sigma^I:=\sigma(e^I) \otimes 1 \otimes 1 \otimes 1$ and the sum on the right hand side is over the representations appearing in a chosen decomposition of the space of quadrilinear invariants. For instance, if we choose $Hom_{\g}(\stackrel{j_0}{\V} \otimes \stackrel{i}{\V} \otimes \stackrel{j}{\V} \otimes \stackrel{k}{\V}, \C) = \bigoplus_s Hom_{\g} (\stackrel{j_0}{\V} \otimes \stackrel{i}{\V}, \stackrel{s}{\V}) \otimes Hom_{\g} ( \stackrel{j}{\V} \otimes \stackrel{k}{\V}, \stackrel{s}{\V^*})$, the four-valent intertwiner appearing in the above equation is given by : 

\beq
\iota_s := \I_{s, j_0 i j k} = \Psi^s_{\h j_0 i} \otimes \Psi_{sjk}.
\eeq

The coefficient appearing in the change of basis between the chosen decomposition and the one given by coupling the representations $j_0-j$ and $i-k$ is given by a six-$j$ symbol.

\begin{figure}[t]
\begin{center}
\includegraphics{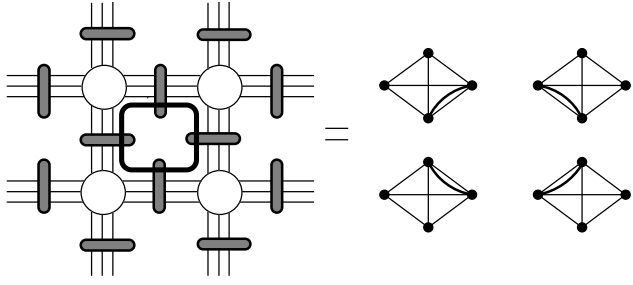}
\caption{Graphical picture of the emergence of the fermionic seven-$j$ symbol.}
\end{center}
\end{figure}
   
Integrating out all group elements and putting everything together, we obtain the following generating functional 

\beq
\label{source3}
\mathcal{Z}^{(I)}(J, \gamma_k, \T) = \prod_{\f} \sum_{j_f} dim(j_f)^{\chi(f)} \left[ \prod_{e \in \gamma_k} \sum_{j_e} \prod_{v \in \gamma_k} \{ 7j (\gamma) \}^I_v (J) \right] \prod_{v \notin \gamma_k} \{ 6j \}_v (J).
\eeq
\\ 
Here, $\{ 6j \}_v (J)$ is the source six-$j$ symbol defined by calculating the usual $\{6j\}$ symbol out of the six representations associated to the six wedges sharing the vertex $v$ with the insertion of the group elements $e^J$. For a given vertex $v$ belonging to six wedges coloured by the spins $j_1,...,j_6$, with associated sources $J_1,...,J_6$ and for a given orientation configuration of the wedges ;

\beqa
\label{6j}
\{ 6j \} (J) &=& \left\lbrace \begin{array}{lll}
j_{1} & j_{2} & j_{3} \\
j_{4} & j_{5} & j_{6} 
\end{array} \right\rbrace(J) \\ \nn
&=& \left( \begin{array}{lll}
a_1 & a_2 & a_3 \\
j_1 & j_2 & j_3 
\end{array} \right)
\left( \begin{array}{lll}
j_1 & a_5 & j_6 \\
a_1' & j_5 & a_6'  
\end{array} \right)
\left( \begin{array}{lll}
j_4 & j_2 & a_6 \\ 
a_4' & a_2' & j_6 
\end{array} \right)
\left( \begin{array}{lll}
a_4 & j_5 & j_3 \\
j_4 & a_5' & a_3'  
\end{array} \right) \times \\ \nn
&& \stackrel{j_1}{\pi}(e^{J_1})_{a_1}^{a_1'} \stackrel{j_2}{\pi}(e^{J_2})_{a_2}^ {a_2'} \stackrel{j_3}{\pi}(e^{J_3})_{a_3}^{a_3'} \stackrel{j_4}{\pi}(e^{J_4})_{a_4}^{a_4'} \stackrel{j_5}{\pi}(e^{J_5})_{a_5}^{ a_5'} \stackrel{j_6}{\pi}(e^{J_6})_{a_6}^{a_6'} .
\eeqa
\\
The six-$j$ symbol (with source insertions) can be conveniently rewritten as a graph reproducing the tensor contraction pattern. This graph is built out of four nodes $n$ and six links $l$ respectively coloured by intertwiners, issued from the Haar integrations, and irreducible representations of $\g$ coming from the colouring of the faces meeting on the vertex $v$. The resulting tetrahedral graph is dual to the original tetrahedron of the triangulation $\T$; his nodes (resp links and triangles) correspond to the triangles (resp segments and vertices) of the original tetrahedron. This means that the three links emerging from a node of the spin network correspond to the three segments building up the corresponding triangle of the original tetrahedron. 
Such a graph defined up to topology and encoding combinatorial representation theory data is called a spin network and can be pictured as follows

\beq
\label{spinnetwork}
\{ 6j \} (J) = \begin{array}{l} \includegraphics[width=2cm,height=2cm]{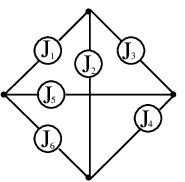} 
               \end{array}.
\eeq
\\
We will fix are conventions such that all spin networks will be built out of normalised intertwiners.

By $\{ 7j(\gamma) \}^I_v (J)$, we mean the modified $\{6j\}$ symbol calculated by taking into account the fermionic line (and its associated sigma matrix) going trough the vertex of interest. Figure $3$ pictures the integration over the group variables yielding the fermionic seven-$j$.
For a given vertex $v$ traversed by a fermionic line and belonging to six wedges coloured by the spins $j_1,...,j_6$ with associated sources $J_1,...,J_6$, the fermionic seven-$j$ yields

\beqa
\label{7j}
\{ 7j \}^I (J) 
&=& \left( \begin{array}{lll}
a_1 & a_2 & a_3 \\
j_1 & j_2 & j_3 
\end{array} \right)
\left( \begin{array}{lll}
j_1 & a_5 & j_6 \\
a_1' & j_5 & a_6'  
\end{array} \right)
\times \\ \nn
&& \left( \begin{array}{llll}
j_0 & j_4 & j_2 & a_6 \\ 
\alpha & a_4' & a_2' & j_6 
\end{array} \right) 
\left( \begin{array}{llll}
\beta & a_4 & j_5 & j_3 \\
j_0 & j_4 & a_5' & a_3'  
\end{array} \right) \h \sigma^{I \alpha}_{\;\;\; \beta} \times \\ \nn
&& \stackrel{j_1}{\pi}(e^{J_1})_{a_1}^{a_1'} \stackrel{j_2}{\pi}(e^{J_2})_{a_2}^ {a_2'} \stackrel{j_3}{\pi}(e^{J_3})_{a_3}^{a_3'} \stackrel{j_4}{\pi}(e^{J_4})_{a_4}^{a_4'} \stackrel{j_5}{\pi}(e^{J_5})_{a_5}^{ a_5'} \stackrel{j_6}{\pi}(e^{J_6})_{a_6}^{a_6'} ,
\eeqa
\\
where $\iota_{j_0 ijk} (\overset{j_0}{e}_{\alpha} \otimes \overset{i}{e}_a \otimes \overset{j}{e}_b \otimes \overset{k}{e}_c) := \left( \begin{array}{llll}
j_0 & i & j & k \\ 
\alpha & a & b & c \end{array} \right)$.

\vspace{2mm}

Equivalently, the seven-$j$ is defined graphically by :

\beq
\label{fermion6j}
\{ 7j (\gamma) \}^I (J) = \sqrt{6} \begin{array}{l} \includegraphics[width=2cm,height=2cm]{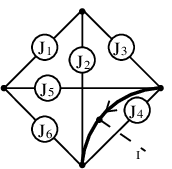} 
                 \end{array}.
\eeq
\\
If $k=2$, the one half line in the fermionic seven-$j$ picture closes by going from one node to itself. 

It is interesting to stress that evaluating the obtained partition function at $J=0$ yields the expectation value of a fermionic Wilson line (with sigma matrices insertions) coupled to three dimensional quantum gravity.

\subsubsection{Grasping operators}

Recall that for a given fermionic loop union $\bar{\gamma}_k=\gamma_k \cup \gamma_0$ of order $k$, with $\gamma_k$ denoting our generic decorated loop defined above, the weight in the total partition function \eqref{partition1} is provided by \eqref{model}, with a by now precise definition of $\mathcal{Z}^{(I)}(J, \gamma_k, \T)$. The last step hence consists in understanding the action of the differential operators acting on the generating functional. The key point is the following. If $J$ is a $\mathfrak{\gh}$-valued current, $J=J^I X_I$, then the action of source derivation on an exponentiated current in the representation $j$ yields  

\beq
\frac{\delta}{\delta J_I} \Pj(e^{J}) \left. \right\vert_{J=0} = \hspace{1mm} \stackrel{j}{\pi}_*(X^I) ,
\eeq
\\
i.e the image of the $I$-th generator of a given (dual) basis of $\hat{\mathfrak{g}}$ under the representation $j$ of the Lie algebra. The crucial point relies on the fact that the Lie algebra homomorphism $\stackrel{j}{\pi}_* : \hat{\mathfrak{g}} \rightarrow \h \Vj \otimes \stackrel{j}{\V^*}$ is an intertwining operator between the adjoint representation, a $j$ representation and its dual.
We note $\Phi$ the associated normalised intertwiner;

\beq
\stackrel{j}{\pi}_* = \Theta(j) \h \Phi^{jj}_{\h 1},
\eeq
\\
where $\Theta(j)^2:=\Theta(1,j,j)^2=<\hspace{1mm} \stackrel{j}{\pi}_*,\hspace{1mm} \stackrel{j}{\pi}_*> = j(j+1)(2j+1)$.

Therefore the action of source derivation with respect to the $I$-th component of the current, also called grasping, modifies a spin network with source insertion by creating a new vertex attached to an open line coloured by the index $I$ in the adjoint representation. From now on, we will represent a link in the adjoint representation by a dashed line:

\beq
\frac{\delta}{\delta J_I} 
\left(  \begin{array}{l}
\includegraphics[width=.5cm,height=1.5cm]{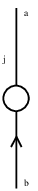}
        \end{array} \right) \left. \right\vert_{J=0} 
	= \Theta(j)
\left( \begin{array}{l} \includegraphics[width=1.5cm,height=1.5cm]{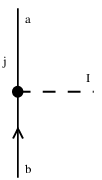} 
       \end{array} \right).
\eeq
\\

\paragraph{The no fermion case.} For all fermionic loop ensembles $\bar{\gamma}_k$ of the form $\bar{\gamma}_p \cup \bar{\gamma}_q$, where $p$ and $q$ are smaller than $n$, there is a squared volume living on each vertex where no fermions are present. This is a remnant of the presence of the mass term in the Dirac action. The squared volume of the three-simplex associated to a vertex $v \in \T^*$ is calculated by acting twice with the volume operator 

\beq
\hat{V}_v = - \frac{i}{8 . 16 . 3!} \sum_{w,w',w'' \supset v} \epsilon_{IJK} 
\frac{\delta}{\delta J_{wI}} \frac{\delta}{\delta J_{w'J}} \frac{\delta}{\delta J_{w''K}} sgn(w,w',w''),
\eeq
\\
on the six-$j$ symbol with source insertions $\{6j\}_v(J)$ \eqref{spinnetwork} living on $v$ and by evaluating the result at $J=0$.
The action of $\hat{V}_v$ associated to a given triple of wedges will create one intertwiner on each of the links $l,l',l''$ corresponding to the wedges $w,w',w''$ mapping the dual adjoint representation space into the second tensor power of the representation space associated to the link. The three adjoint representations spaces then get mapped into the complex via the totally antisymmetric three dimensional Levi-Civita tensor regarded as three-$j$ symbol, i.e. an evaluated three valent intertwiner:

\beq
\epsilon(e_I \otimes e_J \otimes e_K):= \frac{1}{\sqrt{6}}\epsilon_{IJK} = \left( \begin{array}{lll}
1 & 1 & 1 \\ 
I & J & K 
\end{array} \right).
\eeq
\\
Graphically, the result is therefore an ordinary six-$j$ symbol tetrahedral spin network with an additional internal vertex, coloured by the evaluated intertwiner $\epsilon_{IJK}$, whose legs, living in the adjoint representation, are attached to the three links of the spin network grasped by the source derivations corresponding to the wedges of the derived complex $\T^+$, i.e to the segments of $\T$.
The full contribution of the quantum volume of a given vertex $v$ accordingly results in the following sum 

\beq
V_v = \sum_{grasp} K
\left(	\begin{array}{l}
\includegraphics[width=2cm,height=2cm]{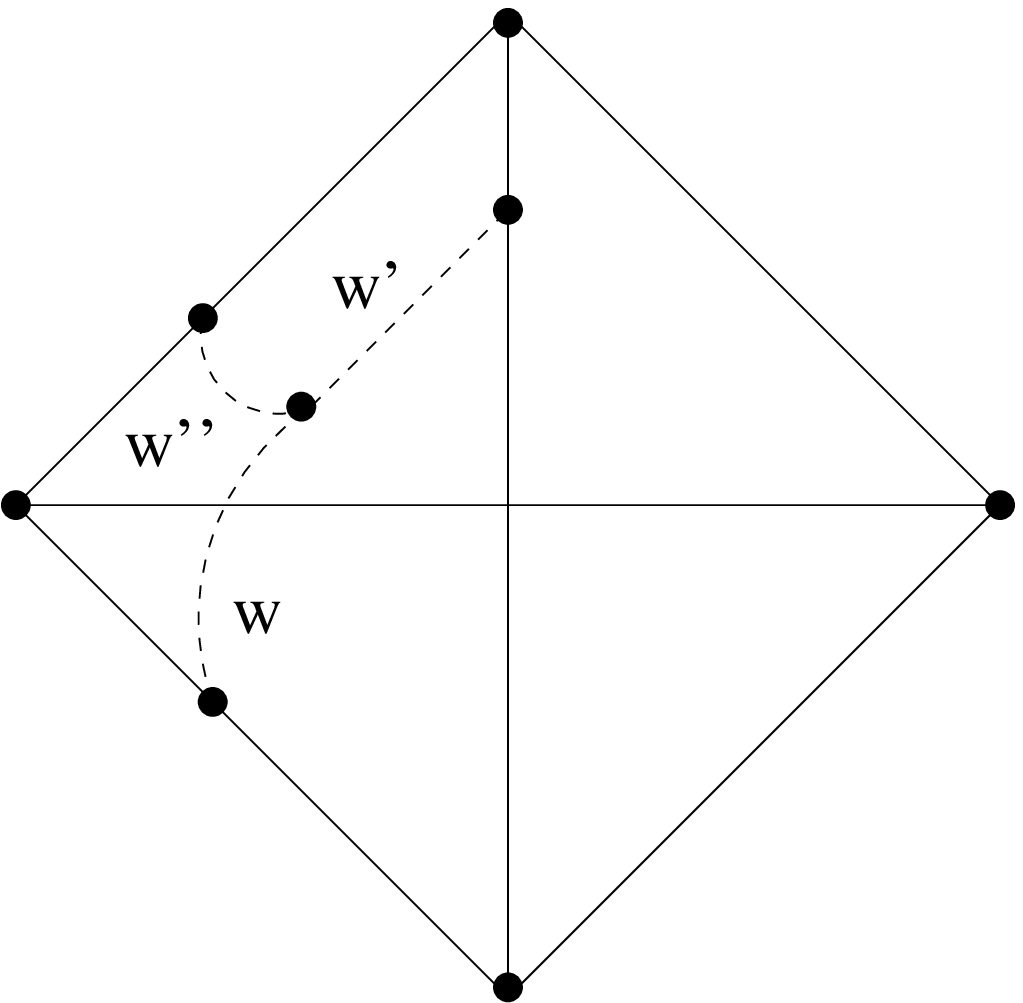}
        \end{array} \right),
\eeq
\\
over all sixteen (plus the permutations) possible graspings on the triple of links of the spin network which correspond the triple of wedges whose dual segments classically span the volume of the tetrahedron dual to the vertex $v$, i.e over all admissible triples.

Here, we have introduced a notational simplification that we will use from now on. The formal sum and the coefficient $K$ above are entirely defined by the spin network at their right. The $\sum_{grasp}$ symbol means that we are summing over all admissible three-graspings of the type defined by the spin network. The first term in the sum is the one pictured. For each such term, the coefficient $K$ is a function of the spins associated to the grasped faces together with the correct sign factor and numerical coefficient. 

Explicitly, for the above case, we have 
  
\beq
\sum_{grasp} : \equiv \sum_{w,w',w''} ,
\eeq
and 
\beqa 
K :&=& K (j_{w},j_{w'};j_{w''}) \\ \nn
&=& - \frac{i}{16.8 \sqrt{6}} sgn(w,w',w'') \Theta(j_{w}) \Theta(j_{w'}) \Theta(j_{w''}) .
\eeqa
\\
The computation of the squared volume $V_v^2$ of a given vertex $v$ now involves the repeated action of $V_v$ on the six-$j$ symbol with source insertions $\{6j\}_v(J)$ \eqref{spinnetwork} associated to $v$ before evaluation at $J=0$. The result, in terms of spin networks, is the usual tetrahedral six-$j$ spin network with now the insertion of a double trivalent grasping. 

This procedure leads us to a quantisation ambiguity; if at least one of the links belonging to the triple of the first three-grasping coincides with one of the links of the second, which is automatically satisfied, we have to decide wether the action of the second power of the volume grasps ``above" or ``underneath" the grasping of the first volume action. The two configurations yield a different spin network and hence a different number for the quantum measurement of the associated tetrahedron's squared volume. 
To circumvent this difficulty, we propose a natural prescription consisting in summing over both contributions with equal weight factor, i.e we prescribe a symmetrized combination of the two possibilities. Concretely, this means that when we calculate the squared volume, 
we first sum over all possible admissible three-graspings (graspings of the triple of wedges whose dual segments do not belong to the same triangle). As above, there are $16 \times 3!$ such terms. For each fixed contribution, we then sum over all the admissible $16$ (without counting the permutations) second three-grasping possibilities. Out of these terms, there is one where the two three-graspings occur on the same triple of links, six where two out of the three links are grasped twice and nine contributions where only one link is grasped two times. Our prescription is then such that when a term with $n$ links grasped twice occurs in the summation, a sum over the $2^n$ possibilities weighted by a factor $1/2^n$ is implemented.
Everytime that such an ambiguity occurs, the formal notational symbol $\sum_{grasp}$ will contain the implementation of the symmetrization procedure.
As a result,

\beq
\label{grasping1}
V^2_v = \sum_{grasp} K \left( \begin{array}{l}
\includegraphics[width=2cm,height=2cm]{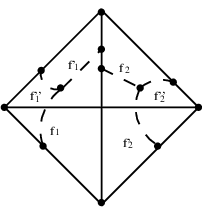}
        \end{array} \right),
\eeq
\\
for each vertex where no fermion is present.
Here,
\beq
\sum_{grasp} :\equiv Sym \left( \sum_{w_{2},w_{2}',w_{2}''} \right) \sum_{w_{1},w_{1}',w_{1}''},
\eeq
and 
\beqa 
K :&=& K_v(j_{w_{1}},j_{w_{1}'},j_{w_{1}''};j_{w_{2}},j_{w_{2}'},j_{w_{2}''}) \\ \nn
&=& - \frac{1}{(16.8)^26} \prod_{i=1}^2 sgn(w_{i},w_{i}',w_{i}'') \Theta(j_{w_i}) \Theta(j_{w_i'}) \Theta(j_{w_i''}),
\eeqa
\\
where the symmetrization is implied by the symbol $Sym(\sum)$ in the second sum.

\paragraph{Fermion loops.} 

We conclude this section by describing the situation along our generic fermionic path $\gamma_k$. Along his travel round the loop $\gamma_k$, the fermion sees a volume contribution $\hat{V}_v$ living on each vertex that he goes trough together with an (internal) area one-form $\hat{A}_{Ie}$ measuring the area of each triangle he is travelling through. The component $I$ of the area one-form measured in the frame associated to the vertex $v$ is calculated by acting with the differential operator

\beq
\hat{A}_{I \hspace{.5mm} uv} = - \frac{1}{12} \sum_{w,w' \supset uv} \epsilon_{IJK} \frac{\delta}{\delta J_{Jw}} \frac{\delta}{\delta J_{Kw'}} sgn(w,w'),
\eeq
\\
on the fermionic seven-$j$ symbol with source insertions $\{ 7j (\gamma) \}^I_v (J)$ \eqref{fermion6j} living on $v$. Graphically, the result on a given pair of links corresponding to a pair of wedges is the creation of two open lines coloured by indices $J$ and $K$ in the adjoint representation. These two lines, together with the open line coloured by the index $I$ coming from $\{ 7j(\gamma) \}^I_v (J)$, finally contract trough the $\epsilon_{IJK}$ intertwining. This is the explicit picture of the soldering of the spin space to the quantum geometry of spacetime.

To complete the picture we also need a volume contribution on the vertex $v$. Here again the procedure implies a double grasping of the fermionic seven-$j$ assigned to the vertices of the fermion loop. The result for a given vertex $u$ is the following

\beqa
\label{grasping2}
\hat{V}_u \hat{A}_{I \hspace{.5mm} uv} \sqrt{6}
\left( \begin{array}{l}
\includegraphics[width=2cm,height=2cm]{7j.eps}
        \end{array}
\right) \left. \right\vert_{J=0}
&=& \sum_{grasp} K
\left(	\begin{array}{l}
\includegraphics[width=2cm,height=2cm]{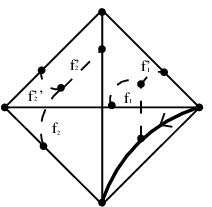}
        \end{array} \right).
\eeqa
\\
Here, the formal notation means
\beq
\sum_{grasp} :\equiv Sym \left( \sum_{w_{2},w_{2}',w_{2}''} \right) \sum_{w_{1},w_{1}'},
\eeq
and
\beqa
K:&=& K_{u}(j_{w_{1}},j_{w_{1}'};j_{w_{2}},j_{w_{2}'},j_{w_{2}''}) \\ \nn
&=&  \frac{i}{16^2 \sqrt{6}} sgn(w_{1},w_{1}') sgn(w_{2},w_{2}',w_{2}'') \Theta (j_{w_{1}}) \Theta (j_{w_{1'}}) \Theta (j_{w_{2}}) \Theta (j_{w_{2'}}) \Theta (j_{w_{2''}}).
\eeqa
\\
The sum is first over the three possible two-graspings of the links corresponding to the wedges dual to the segments building up the triangle through which the fermion travels. For each such terms there is then a sum over the admissible three-graspings. The $Sym$ symbol ensures that we are symmetrizing the appearing ambiguities. 

We are now ready to give a precise definition of the model since all quantities appearing in \eqref{model} have been defined. 

\subsubsection{The model}

Let us summarise the results and discuss the properties of the model.
The proposal made in this paper for the description of massive fermionic fields coupled to three dimensional quantum gravity consists in an order by order calculation of the partition function \eqref{partition} of the theory through an inverse mass expansion \eqref{partition1}. 

The weight of the decorated loop union $\bar{\gamma}_k=\bar{\gamma}_p \cup \bar{\gamma}_q$, $p+q=k$, of order $k \in [0,2n]$ appearing in the $\alpha$ expansion, is given by the following general expression: 

\beq
\label{model2}
\mathcal{Z}(\bar{\gamma}_k, \T) = \hspace{1mm} \prod_{\f} \sum_{j_f} dim(j_f)^{\chi(f)} \left[ \prod_{e \in \gamma_k} \sum_{j_e} \prod_{v \in \gamma_k} A_{v}(\bar{\gamma}) \right] 
\prod_{v \notin \gamma_k} B_{v}(\bar{\gamma}) .
\eeq
\\
The no-fermion vertex amplitude $B_v$ is non trivial if and only if the loop ensemble $\bar{\gamma}_k$ is built out of the union of two loop configurations $\bar{\gamma}_p$ and $\bar{\gamma}_q$ which are such that $p$ and $q$ are smaller than $n$ (if one of the two loop configurations saturates all vertices, there is obviously no no-fermion situation). In this case, the amplitude is defined to be the square volume \eqref{grasping1} : $B_v(\bar{\gamma}) = V_v^2$.

The vertex amplitude $A_{v}(\gamma)$ sitting on the fermion path depends on the type of loop configuration, decoration considered and also generically varies from one vertex to another. 

For the particular loop configuration $\bar{\gamma}_k = \gamma_k \cup \gamma_0$ that we have used to build the model, the fermionic vertex amplitude has been defined. How does this amplitude generalise to the other contributions ? For instance, how to build amplitudes for other types of configurations and other types of decorations ? 

We have seen how all the other terms in the sum over loop decorations \eqref{sumconfigurations}, i.e all possible decorated loops, generated for a fixed set of $k$ vertices could be obtained and we have defined a classification of inequivalent loop decorations. 
Let us pick one general decorated $k$-loop $\gamma_N \in \mathcal{L}_k$ (more precisely, a chosen representative of an equivalence class in $\mathcal{L}_k / \sim$) whose Dirac trace is built out of $N$ sequences of the form $\sigma U \sigma U ... \sigma U U^{\dagger} \sigma U^{\dagger} \sigma ... U^{\dagger} \sigma$ out of our previously established classification. Call $k-m$ (resp $m$) the total number of type $a$ : $\sigma U$ (resp type $b$: $U^{\dagger} \sigma$) like patterns appearing in $\gamma_N$.
Our generic loop hence corresponds to a decorated loop of the form $\gamma_1$ with $m=0$.
Let $\ve_{\gamma_N} = \ve \cap \gamma_N$ denote the $k$ vertices traversed by the fermionic loop and $\ve_a$ (resp $\ve_b$) the subset of vertices associated to the portions of the loop corresponding to the type $a$ (resp type $b$) sub-sequences  which are not a starting point for a type $b$ (resp type $a$) sub-sequence.
We call $\mathcal{V}_{ab}$ and $\mathcal{V}_{ba}$ the subset of vertices of $\ve_{\gamma_N}$ that respectively correspond to an end point of a sub-sequence of type $a$ and to an end point of sub-sequence of type $b$. Hence, $\ve_{\gamma_N} = \ve_a \cup \mathcal{V}_{ab} \cup \ve_b \cup \mathcal{V}_{ba}$.

Taking products of such decorated loops $\gamma$, we obtain loop configurations $\bar{\gamma}$.
The fermionic determinant generates quadratic loop configurations of the form $\bar{\gamma}_k = \bar{\gamma}_p \cup \bar{\gamma}_q$. The value of $p$ and $q$ determines three classes of loop configurations.   

We now show how the weights associated to all the possible contributions appearing in the fermionic loop expansion \eqref{expansion} can be calculated from the forgoing construction by separately considering the three different types of quadratic loop configurations.  

\paragraph{The case $p,q <n$.} If both $p$ and $q$ are strictly smaller than the number of vertices $n$, we generically observe non degenerate vertices. However, the two loop configurations $\bar{\gamma}_p$ and $\bar{\gamma}_q$ can be supported, partially or totally, by a common set of vertices. 
We first suppose that $\bar{\gamma}_p \cap \bar{\gamma}_q = \oslash$ and consider one component $\gamma_N \subset \bar{\gamma}_p$ of order $r \leq p$.

We now follow a fermion along his travel around the loop $\gamma_{N}$ starting from a portion of the loop associated to a type $a$ sub-sequence belonging to a given previously defined sequence $n$ out of the $N$.  
Along all the sub-sequence of type $a$, all vertex amplitudes are given by the following vertex functions. Forall $v \in \ve_a$,

\beq
\label{amplitude1}
A_{v}(\gamma_N)= (-1)^m \sum_{grasp} 
K
\left(	\begin{array}{l}
\includegraphics[width=2cm,height=2cm]{tetrahedron4.eps}
        \end{array} \right)
\eeq
\\
for $k \neq 2$ and 
\beq
\label{amplitude2}
A_{v}(\gamma_1)=\sum_{grasp} 
K 
\left(	\begin{array}{l}
\includegraphics[width=2cm,height=2cm]{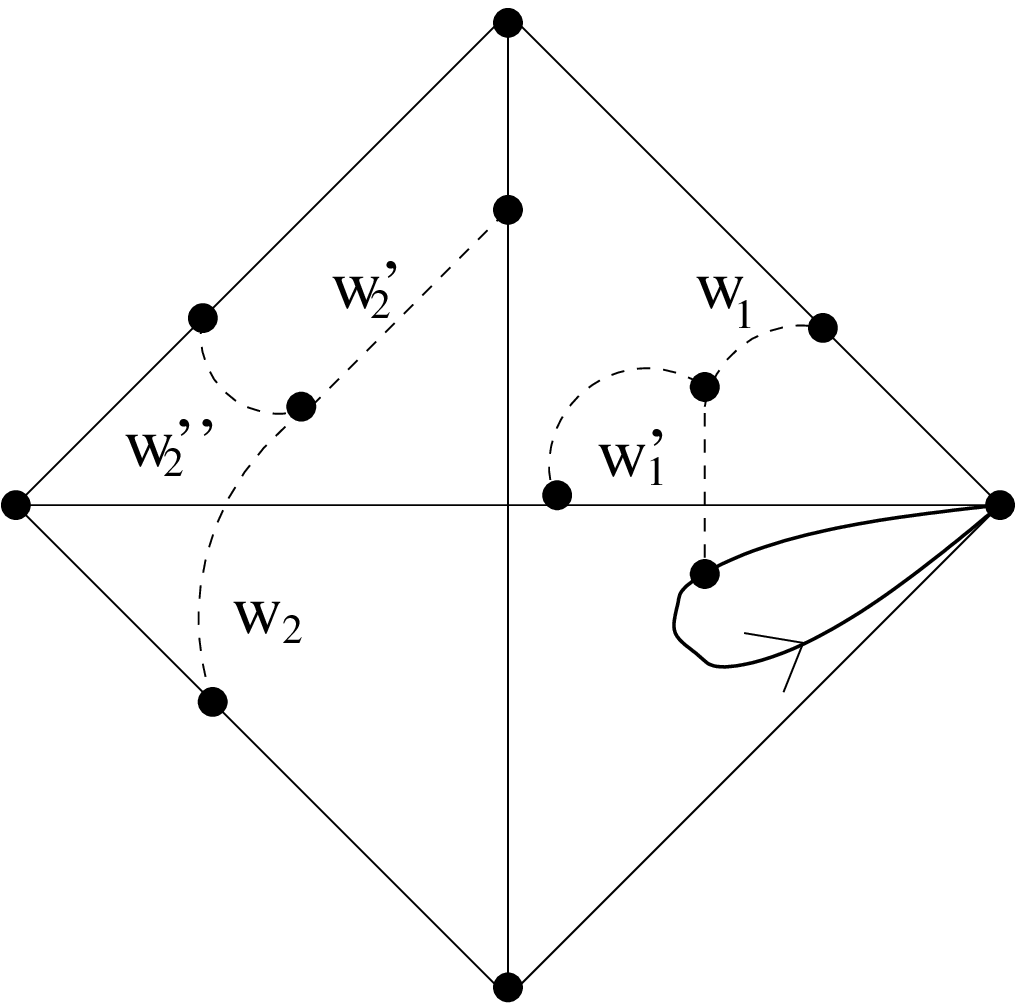}
        \end{array} \right)
\eeq
for the second order case.

The fermion now observes a junction between a sub-sequence of type $a$ and a sub-sequence of type $b$ yielding a pattern in the trace of the form $U U'^{\dagger}$. Accordingly, if $k\neq2$, all vertices $v \in \mathcal{V}_{ab}$ carry the following vertex amplitude

\beq
A_{v}(\gamma_N) = 
(-1)^m\left(	\begin{array}{l}
\includegraphics[width=2cm,height=2cm]{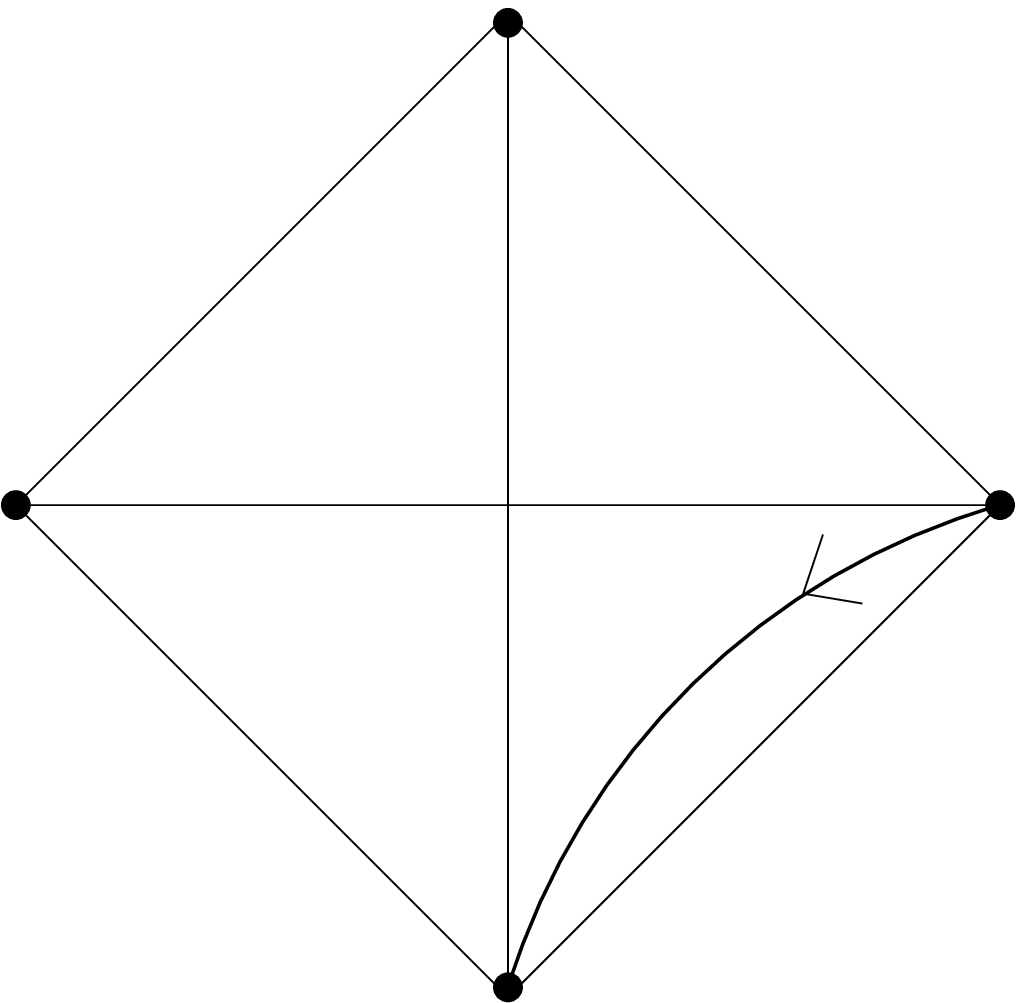}
        \end{array} \right),	
\eeq
\\
while the $k=2$ case yields
\beq
\label{close}
A_{v}(\gamma_1) = \sum_{grasp} 
-2 K 
\left(	\begin{array}{l}
\includegraphics[width=2cm,height=2cm]{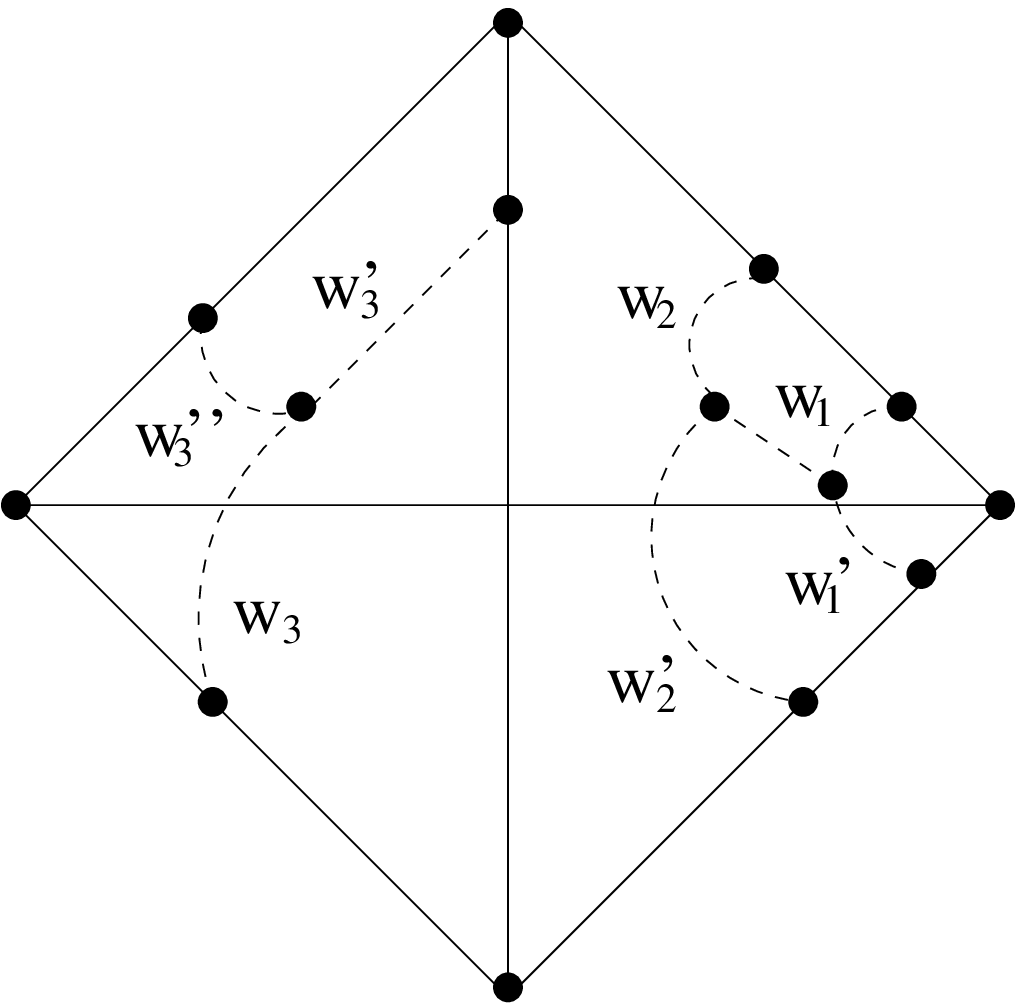}
        \end{array} \right),	
\eeq
\\
because the two group elements $U$ and $U'^{\dagger}$ correspond to the same edge: $U'=U$. The factor two comes from the trace formula $Tr_D(\sigma^I \sigma^J)=2\eta^{IJ}$.
Note that the formal sum over the graspings here contains a symmetrizing prescription for triple graspings of the same link: $\sum_{grasp}=Sym(\sum_{w_3,w'_3,w''_3} Sym (\sum_{w_2,w'_2})) \sum_{w_1,w'_1}$.
In this particular case, the sum over the intertwiners in \eqref{model2} disappears. It is interesting to remark that, omitting the volume contribution, we obtain the expectation value of the area of the triangle dual to the edge on which the fermion is travelling.

Following his trip, the fermion is now going through a type $b$ sub-sequence. All amplitudes for $k \neq 2$ and for vertices $v$ belonging to $\ve_b$ will be of the form 

\beq
A_{v}(\gamma_N) = (-1)^m \sum_{grasp} 
K 
\left(	\begin{array}{l}
\includegraphics[width=2cm,height=2cm]{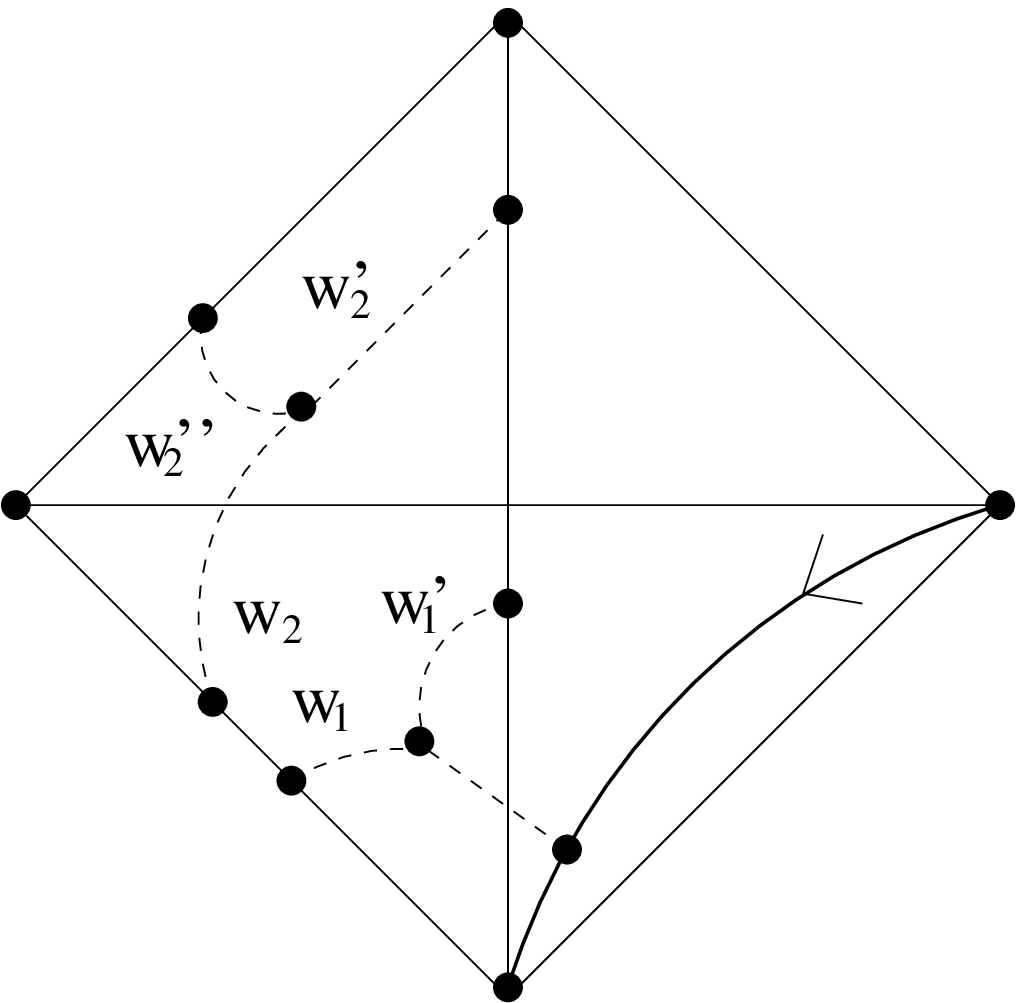}
        \end{array} \right).	
\eeq
\\
Finally, the fermions reaches the boundary of a type $b$ sub-sequence to end the sequence $n$. If $n \neq N$, a new type $a$ sub-sequence belonging to a new $n+1$ sequence starts. Here two, and only two, Pauli matrices are brought together inside the Dirac trace. We accordingly use the relation $\sigma^I \sigma^J = i \epsilon^{IJ}_{\hspace{1.5mm} K} \sigma^K + \eta^{IJ}$,
creating two types of terms associated to the disappearance of the adjacent sigma matrices. 

Using the tools developed in this paper, it is not hard to realise that $\forall v \in \ve_{ba}$, the weights are given by
 
\beq
A_{v}(\gamma_N) = \sum_{grasp} 
(-1)^m K 
\left[ 6i \left(	\begin{array}{l}
\includegraphics[width=2cm,height=2cm]{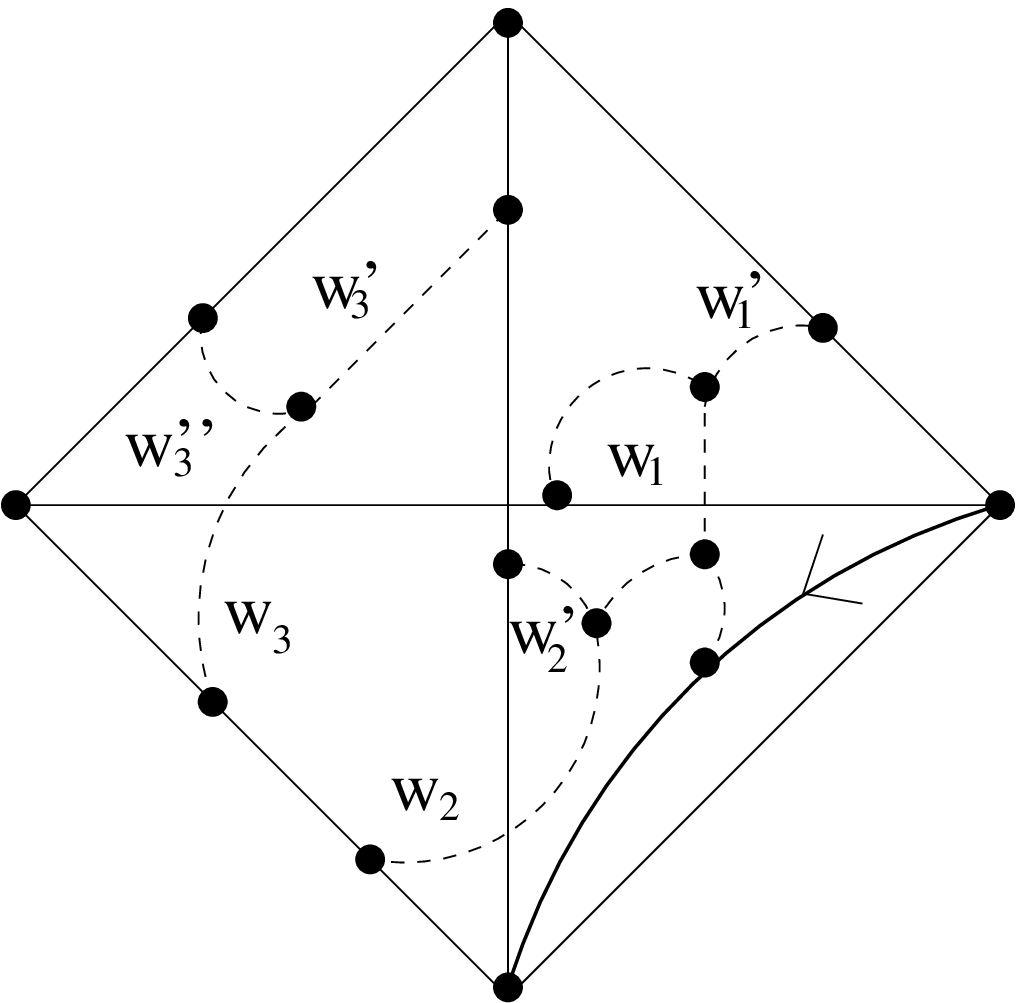}
        \end{array} \right) +
\left(	\begin{array}{l}
\includegraphics[width=2cm,height=2cm]{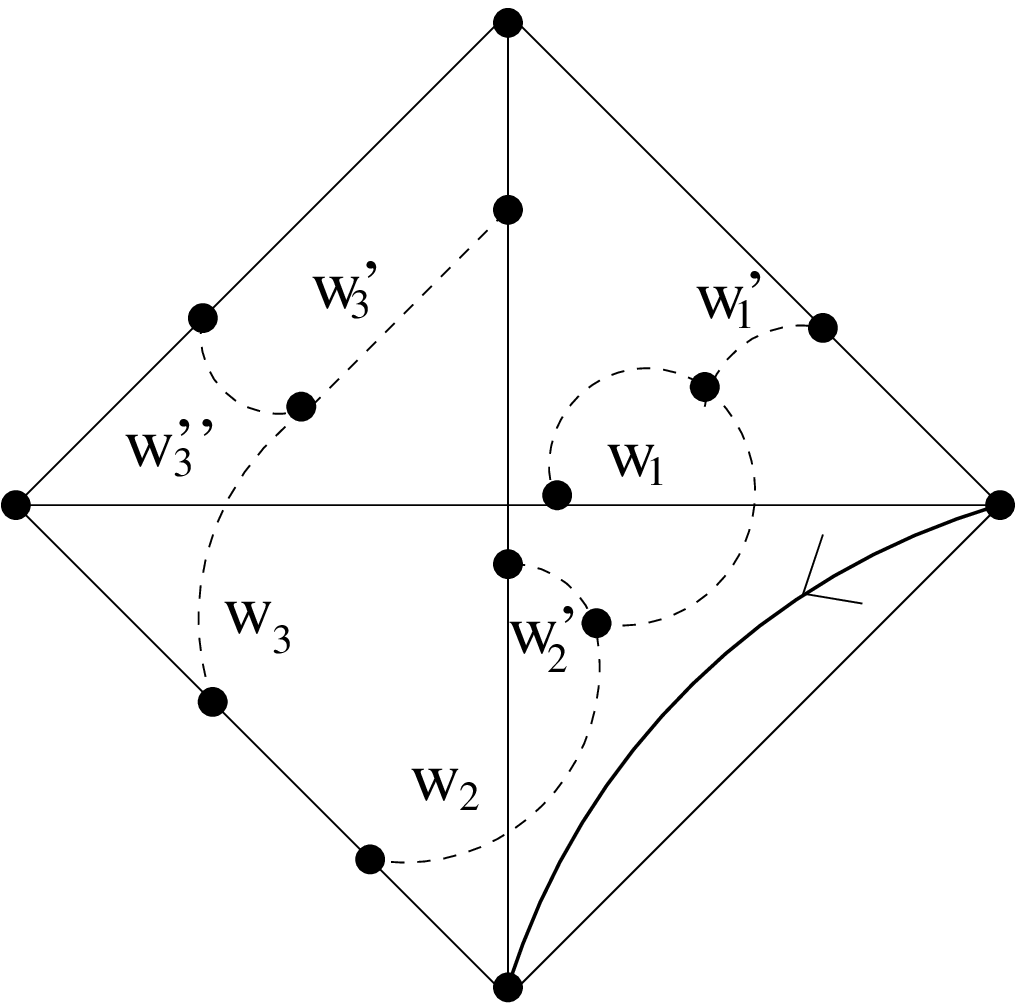}
        \end{array} \right) \right].	
\eeq
\\
Together with the no-fermion vertex amplitude, the above vertex functions define the model completely for non-degenerate loop configuration with $p,q<n$. 

Now, if $\bar{\gamma}_p \cap \bar{\gamma}_q \neq \oslash$, we observe degenerate vertices traversed by two fermionic lines.
If $v \in \ve_a$ is a degenerate vertex, we obtain the following amplitude forall $k \neq 4$, and for a given relative orientation of the two fermionic lines

\beq
A_{v}(\gamma_N) = (-1)^m \sum_{grasp} 
K
\left(	\begin{array}{l}
\includegraphics[width=2cm,height=2cm]{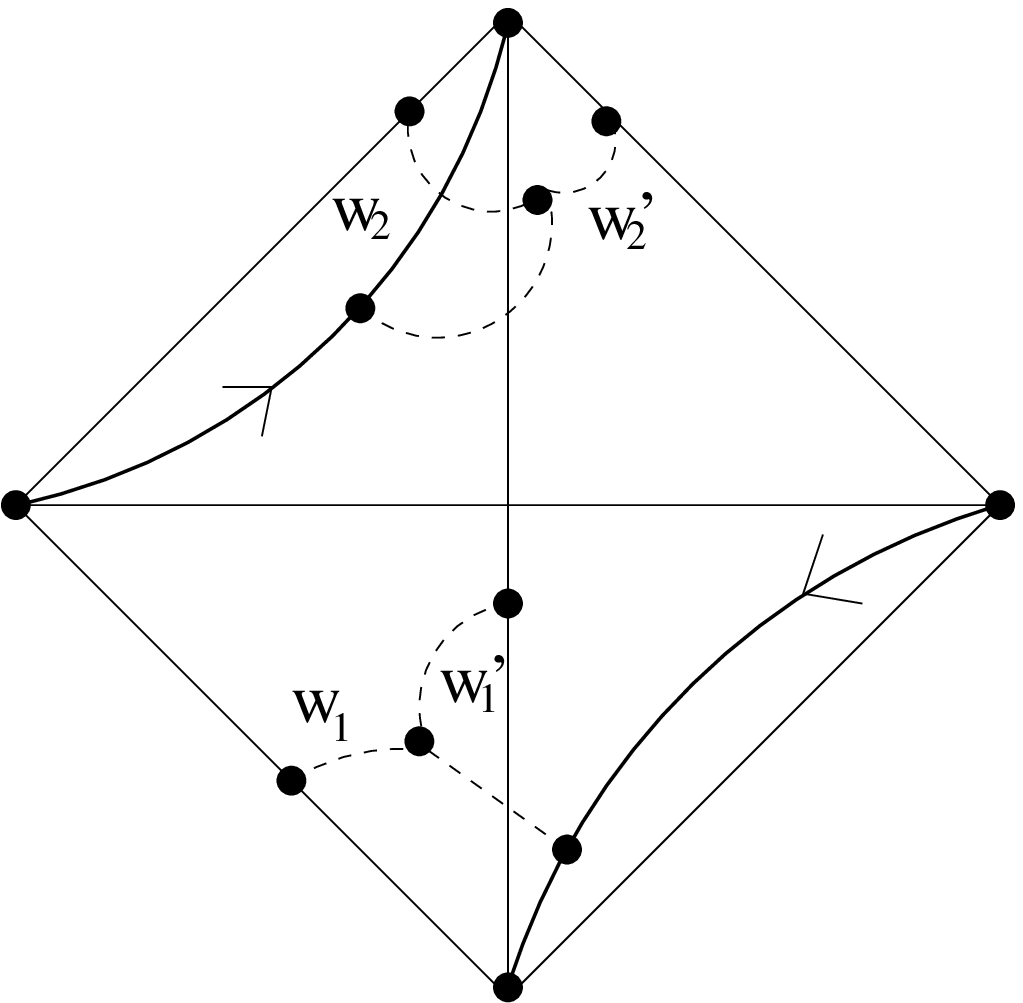}
        \end{array} \right),
\eeq
\\
or
\beq	
A_{v}(\gamma_N) = (-1)^m \sum_{grasp}
K
\left(	\begin{array}{l}
\includegraphics[width=2cm,height=2cm]{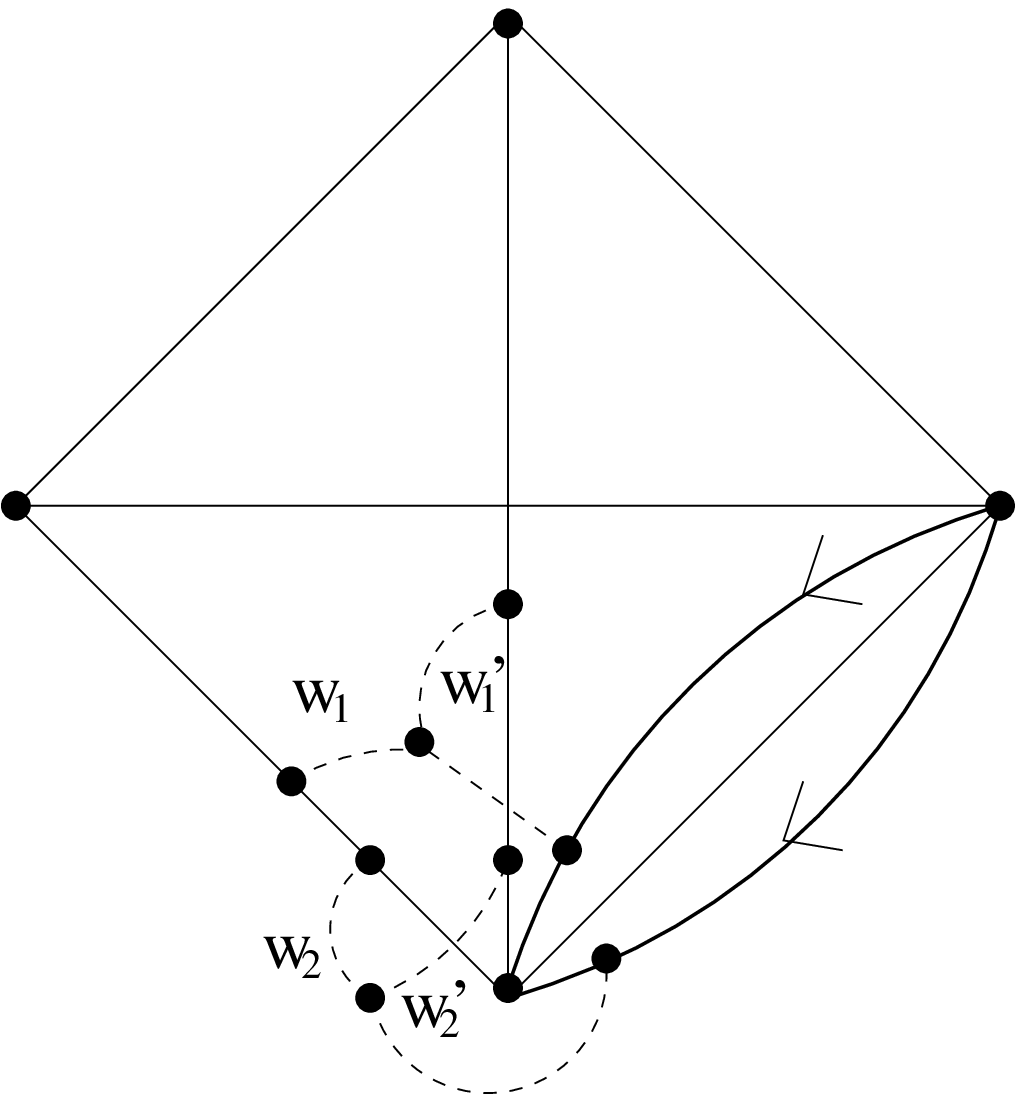}
        \end{array} \right),	
\eeq
\\
depending on the number of fermionic lines living on the same edge converging towards the degenerate vertex. If $k=4$, the two fermionic lines simply close, as in \eqref{amplitude2}.
All these amplitudes trivially generalise to vertices $v \in \ve_{ab}, \ve_b, \ve_{ba}$ by using the results obtained in the non degenerate case. 

\paragraph{The $p,q \leq n$ case.} If $p<n$ and $q=n$, we necessarily obtain degenerate quadratic loop configurations. Some vertices, namely the vertices in the complementary on the $\bar{\gamma}_p$ configuration, however remain non degenerate. If $p=q=n$, all vertices are degenerate. In all these cases, the results developed above generalise yielding the fermionic vertex amplitudes $A_v(\gamma)$. Recalling that the no-fermion $B_v(\gamma)$ amplitudes are trivial, we have now written down all possible vertex amplitudes and the model is consequently defined for all type of loops emerging from the hopping parameter expansion. The sum of all these contributions yields the partition function of the coupled system.
This concludes our presentation of our proposed model of massive fermions coupled to three-dimensional quantum gravity.

\section{Conclusion}

In this paper, we have proposed a model describing the coupling of massive fermions to $3d$ quantum gravity by defining the vacuum to vacuum transition amplitude of the theory. We have first defined a consistent discretization procedure yielding a concrete meaning to the functional measures. Then, we have implemented the Berezin integral over the fermionic fields obtaining the expectation value of a functional determinant. This determinant has then been calculated by using a hopping parameter expansion yielding a sum over fermionic loops. This finite expansion has enabled us to compute the path integral of the theory as we have shown how to calculate the path integral order by order in the inverse mass expansion parameter. The integration over simplicial co-frames is realised by source derivation on a generalised generating functional where the integration over discretized connections is implemented. It is interesting to note that a two-dimensional slicing of one of the contributions to the above spinfoam model associated to a given loop, non co-planar with the slicing, yields a spin network with an open end coloured by the spinor representation, in accordance with loop quantum gravity.
The model is rich enough to take into account the effects of quantum torsion and of non trivial curvature and highlights the subtle relation between spinors and quantum geometry. 

Many issues remain open. First, the emergence of quantisation ambiguities due to the presence of high order polynomials in the simplicial triad field in the theory is clearly a problem. We have prescribed a coherent procedure but one would need to show that the result does not depend on the prescription used to complete the picture.

Second, unlike the point particle case, fields break the topological invariance of three dimensional gravity. The exactitude of the model is hence questionable, even if the spinfoam explicitly derives from a classical action. This leads to two questions. First, do the asymptotics ($j \rightarrow \infty$) of the vertex amplitudes reproduce (a simple function of) the classical Regge action coupled to fermions ? Second, and most importantly : what is the behaviour of the model under refinement of the triangulation? Since the finiteness of our loop expansion is due to the cut-off on  the number of degrees of freedom induced by the triangulation, this issue is of primary interest, especially because of the large value of the expansion parameter. 
An alternative to these issues is to consider the GFT approach \cite{GFT3} to implement a sum over topologies and triangulations needed when one works with non-topological theories. 

Third, the model obviously diverges in the infrared as for pure (non properly gauged-fixed \cite{Freidel}) gravity because of the infinite sum over representations. A possible regularization scheme to explore consists in incorporating a positive cosmological constant $\Lambda$ in the theory leading to a modified Turaev-Viro model \cite{Turaev} based on the representation theory of the real compact Hopf algebra $\mathfrak{U}_q(su(2))$ with the deformation parameter $q$ at root of unity related to the cosmological constant $\Lambda$ via $q=e^{i l_p \sqrt{\Lambda}}$. Since there are only a finite number of representations of $\mathfrak{U}_q(su(2))$ (with non-vanishing $q$-dimension) at root of unity, this procedure would cut of the infinite sum over representations and yield an infrared finite model. 

Finally, one would need to analyse the well known fermion doubling problem appearing in lattice-like theories incorporating fermions. Does the spinfoam generate doublers ? Could we modify the action to suppress the extra species ?

It is interesting to underline the crucial differences between the coupling of fields and point particle to quantum gravity. First, the effects of fields on geometry are non-localised, in the sense that matter fields create non trivial vertex amplitudes (squared volumes for instance) on all vertices of the dual complex. These effects are more difficult to control and more complicated than in the point particle case where curvature and torsion are non trivial only around the particle's worldline (note however that this could be due to the mass term). Second, as discussed above, fields break the topological invariance of three dimensional gravity for the same reasons, whereas particles do not.

To conclude, we discuss the perspectives. The first natural extention seems to be the generalisation to four spacetime dimensions. At first, one might expect the coupling of fermions to the Barrett-Crane model of four dimensional quantum gravity to be problematic. As remarked in \cite{Miko3}, the reason is that the kinetic term of the standard Dirac action in a $d$-dimensional curved spacetime couples to the $d-1$ exterior power of the $d$-bein. Hence, in even dimensions, fermions are coupled to Palatini gravity via an odd power of the gravitational field not convertible into any power of the $d-2$-form B field of BF theory. However, if one uses the Capovilla, Dell, Jacobson, Mason chiral action \cite{Capovilla} of fermions coupled to two-form gravity, the fermions indeed couple to the B field and one could imagine such an action as an interesting  starting point for a four dimensional construction \footnote{We thank the referees for pointing out this possibility as a potential extension of the work presented here.}.
Next, let us comment on the Lorentzian framework. Because of the non-compacity of the associated symmetry group, the unitary, irreducible representations, appearing in the character expansion of the delta function forcing the discretized curvature to be trivial around the faces of the spinfoam, would be infinite dimensional. Since the fermions live in finite dimensional representations, we may expect a problem in the integration over the discretized connections since the space of non-vanishing intertwiners between the finite and the unitary representations is empty. Hence, a generalisation of the present work to Lorentzian signatures seems a priori non-trivial.
Finally, inspired by the recent series of papers devoted to the computation of the graviton propagator \cite{graviton}, it would be interesting to compute the two point function of the model. The techniques developed in this paper should be easily generalisable to the construction of the fermion propagator in three-dimensional spinfoam quantum gravity. 
 
\subsection*{Acknowledgements}

I would like to thank Laurent Lellouch for informing me about the hopping parameter expansion technique in lattice QCD, Philippe Roche for reading the manuscript, Alejandro Perez for a non-countable set of exciting discussions and important suggestions, John Barrett for a stimulating fermionic dinner. Finally, and especially, I am indebted to Carlo Rovelli for his support, encouragements and his deep crucial insights on some key aspects of this paper.

\appendix

\section{Appendix : Geometrical and algebraic setting} 

This appendix is devoted to review the geometrical and algebraic structures underlying the coupling of fermions to a three dimensional dynamical spacetime. 
We first define the Clifford algebra of interest in this paper, its associated spin group and spinors before introducing the notion of spinor bundles and spin connections.
Most of the results reviewed here are inspired by references \cite{geometry}.

\subsection{Fermions in flat spacetime}

Flat Euclidean spacetime spinors are related to,

\subsubsection{The Clifford algebra $\mathcal{C}(3)$ and the spin group $Spin(3)$} 

The Clifford algebra $\mathcal{C}(3,0) \equiv \mathcal{C}(3)$ is the real associative algebra generated by a unit $\mathbb{I}$ and the three symbols $\gamma^1, \gamma^2, \gamma^3$ satisfying

\begin{align}
\gamma^I \gamma^J + \gamma^J \gamma^I = 2 \eta^{IJ}, \hspace{2mm} \mbox{where} 
\hspace{2mm} 
\eta^{IJ}=\begin{cases}
              1 & \textrm{if $I=J=1,2,3$}  \\
              0 & \textrm{if $I \neq J$}.  
           \end{cases}
\end{align} 
\\
A possible vectorial basis of $\mathcal{C}(3)$ is provided by 
$\{\mathbb{I}, \gamma^I, \gamma^I \gamma^J, \gamma^I \gamma^J \gamma^K \}_{I < J < K}$. The Clifford algebra $\mathcal{C}(3)$ is therefore a linear space of dimension eight isomorphic to the exterior algebra $\Lambda(\mathbb{R}^3)$ as a vector space and to the space of two by two complex matrices $M_2(\mathbb{C})$ as a real associative algebra.
The later isomorphism provides us with a faithful representation of $\mathcal{C}(3)$ on $\mathbb{C}^2$:

\beqa
\rho : \mathcal{C}(3) &\rightarrow& End (\mathbb{C}^2) \\ \nonumber
              \gamma^I  &\rightarrow& \rho(\gamma^I)=\sigma^I,
\eeqa
\\
in terms of the Pauli matrices $\sigma^I$, $I=1,2,3$, 

\beq
\sigma^1 = \rho(\gamma^1) =
\left( \begin{array}{ll}
0 & 1 \\ \nn
1 & 0 
       \end{array} \right) ,
\hspace{2mm} 
\sigma^2 = \rho(\gamma^2) =
\left( \begin{array}{ll}
0 & -i \\ \nn
i & 0 
       \end{array} \right) ,     
\hspace{2mm} 
\sigma^3 = \rho(\gamma^3) =
\left( \begin{array}{ll}
1 & 0 \\ \nn
0 & -1 
       \end{array} \right) ,      
\eeq
\\
known as the Pauli algebra. The unit is mapped onto the unit two dimensional matrix.
Let $\mathcal{C}_0 (3)$ be the even subalgebra of $\mathcal{C}(3)$, i.e the subalgebra linearly generated by products of an even number of generators $\gamma$ (i.e two generators) and let $\mathcal{C}_1 (3)$ denote the vector subspace of $\mathcal{C}(3)$ linearly generated by the elements $\{\gamma^I\}_I$.

The group $Spin(3)$ can be explicitly realized as a subset of the Clifford algebra
$\mathcal{C}(3)$. To see this, let $V$ denote a real vector space of dimension three endowed with a non-degenerated bilinear form $g$ and denote by $\{e_I\}_I$ a $g$-orthonormal basis of $V$: $g(e_I,e_J)=\eta_{IJ}$. It is possible to embed the algebraic dual $V^*=\mbox{Span}\{e^I\}_I$ (endowed with the metric with components $\eta^{IJ}$) of $V$ in $\mathcal{C}(3)$ through the linear map
 
\beqa
\label{embedding}
\gamma: V^* &\rightarrow& \mathcal{C}_1 (3) \subset \mathcal{C}(3) \\ \nn
 v=v_I e^I  &\rightarrow& \gamma(v)=v_I \gamma^I:= \vslash.  
\eeqa
\\
Consider the subset of elements $s \in \mathcal{C}(3)$ that are invertible and such that
 
\beq
\forall v \in V^*, \hspace{2mm}  \chi(s)(\gamma(v))= s \gamma(v) s^{-1} \in \mathcal{C}_1 (3).
\eeq
\\ 
The set of such elements $s$ forms a group called the Clifford group $\Gamma$.
Note that the map $\chi(s):\mathcal{C}_1 (3) \rightarrow \mathcal{C}_1 (3)$ provides us with a linear transformation of $V$ because $V$, $V^*$ and  $\gamma(V^*)=\mathcal{C}_1(3)$ are isomorphic as vector spaces. The first isomorphism is due to the fact that the bilinear form $g$ is non-degenerate and the second one is given by \eqref{embedding}.
More precisely, $\chi(s)$ is an isometry of the bilinear form $g$, i.e an orthogonal transformation belonging to $O(3)$. The representation of the Clifford group given by $\chi:\Gamma \rightarrow GL(V^*)$ is not faithful because, for instance, $s$ and $as, a \in \mathbb{R}^*$, yield the same rotation through $\chi$.
A standart way of distinguishing between $s$ and $as$ is to introduce a normalisation.
$Spin(3)$ is then derived from the Clifford group by using this normalisation condition. Denote by a ``bar" the involution of $\mathcal{C}(3)$ defined by $\overline{\gamma^1 \gamma^2...\gamma^p}=\gamma^p...\gamma^2 \gamma^1$. One can now define the spin group as the following subgroup of $\Gamma$

\beq
\label{spingroup}
Spin(3)=\{s \in \Gamma \cap \mathcal{C}_0 (3) \hspace{1.2mm} \mbox{such that} \hspace{1mm} \mid \bar{s} s \mid  = \bar{s} s=1\},
\eeq
\\
where $\mid . \mid$ denotes the absolute value.
One can show that $\chi(Spin(3))=SO(3)$ and that the map $\chi: Spin(3) \rightarrow SO(3)$ is a surjective two to one group morphism which implies that $\hat{G}=Spin(3) \simeq SU(2)$ is the universal double cover of $G=SO(3)$ because $Spin(3)$ is simply connected.
We can equivalently  obtain $Spin(3)$ in its fundamental or defining representation as a two by two complex matrix group through the isomorphism $\rho$. Indeed, in $M_2(\mathbb{C})$, the above definition is equivalent to defining the group $Spin(3)$ as the group of invertible $2 \times 2$ matrices $\rho(s)=A$, such that there exists a $SO(3)$ linear transformation $\Lambda(A): V^* \rightarrow V^*$, satisfying:

\beq
\label{spin}
A \hspace{.5mm} \sigma(e^I) \hspace{.5mm} A^{-1} = \sigma(\Lambda^I_{\hspace{1mm} J}(A) e^J) , \hspace{2mm} \mbox{and} \hspace{1mm} det \hspace{1mm} A =1,
\eeq
\\
where $\sigma=\rho \circ \gamma:V^* \rightarrow M_2(\mathbb{C})$ and the constraint on the determinant forces the element $s \in \Gamma$ to belong to the pair subalgebra $\mathcal{C}_0 (3)$ and to satisfy the normalisation condition. We here recognise the fact that the adjoint action of $SU(2)$ on its Lie algebra $\mathfrak{su}(2) \simeq \mathbb{R}^3$ is a rotation of $\mathbb{R}^3$.
Physically, we can read this equation as stating that an orthogonal transformation in spacetime $V$ must always be accompanied by a change of basis in spin space $\mathbb{V}=\mathbb{C}^2$ for the spin rotation $\sigma(v), v \in V$, to be independent of the frame in which the vector $v$ is expressed. This equation also contains information about the profound difference between spinors and vectors, namely that spinors are not invariant under transformations corresponding to non trivial elements of the center of $SU(2)$. Indeed, note that a $2\pi$ radian rotation changes the sign of the spin matrix $A$ of equation \eqref{spin} keeping the associated spacetime rotation unchanged. Therefore a spinor on which the spin matrix acts will feel the difference between a $2\pi$ and a $4\pi$ rotation whereas a vector will not.

\subsubsection{Spinors}

We can now define a fermion. Essentially, a fermion $\psi$ is a physical field whose local excitations yield half integer spin particles and whose dynamics are governed by the Dirac Lagrangian. 
This Lagrangian is built out of complex linear combinations of the fields. Hence, the mathematical definition of a classical spin one-half fermion relies on complexified Clifford algebras.
  
A Dirac spinor is an element of a complex vector space $\mathbb{V}$ on which the complexified even Clifford subalgebra $\mathcal{C}_0 (3)^{\mathbb{C}} = \mathcal{C}_0 (3) \otimes_{\mathbb{R}} \mathbb{C}$ is irreducibly represented. Because of the isomorphisms of complex algebras $\mathcal{C}_0 (3)^{\mathbb{C}} \simeq \mathbb{H}^{\mathbb{C}} \simeq M_2(\mathbb{C})$, where $\mathbb{H}$ denotes the quaternions, we see that this vector space, also called Clifford module, is given by $\mathbb{V}=\mathbb{C}^2$. 
Weyl spinors do not exist here because $\mathcal{C}_0 (3)^{\mathbb{C}}$ is simple: the algebra cannot be broken into a left and a right part by the use of the chirality operator $\sigma^5=i\sigma^1 \sigma^2 \sigma^3$. Moreover, (true) Majorana spinors do not exist either because the representation $\mathbb{V}$ is not of real type. In fact, the representation $\V$ is quaternionic. This means that 
$\mathbb{V}$ can be endowed with a quaternionic structure, i.e a $\C$-linear 
\footnote{We could also define a quaternionic structure as an anti-linear, anti-involutive map $C: \V \rightarrow \V$. The two definitions are equivalent.}
function $C : \V \rightarrow \overline{\V}$, mapping $\V$ onto its complex conjugate vector space $\overline{\V}$, satisfying $\overline{C} \circ C = -id_{\V}$, where $\overline{C} : \overline{\V} \rightarrow \V$ denotes the complex conjugate map to $C$ \cite{trautman}. This structure allows us to define a modified version of Majorana spinors called symplectic Majorana fermions.

Because we have the inclusion \eqref{spingroup} $Spin(3) \subset \mathcal{C}_0 (3)$, the Dirac spinors $\psi \in \mathbb{V}$ also yield a representation of $Spin(3)$. In fact, this representation $\V$ is the fundamental $j=1/2$ representation of $Spin(3)$, where $j \in \mathbb{N}/2$ denotes the spin labelling the irreducible, unitary representations of the compact group. Let us introduce some notation used troughout the paper.
If $\stackrel{j}{\mathbb{V}} = \mathbb{C}^{2j +1} = \C \{ \stackrel{j}{e}_a \h \mid a=-j,...,j \}$, denotes the left $Spin(3)$-module associated to the representation $j$, we have $\stackrel{1/2}{\mathbb{V}}:=\V$.
We will note $\overset{j}{\mathbb{V}}\,^*$ the contragredient representation space and $\{\overset{j}{e}\,^a\}_a$ the basis dual to $\{\stackrel{j}{e}_a\}_a$.
We will call $\stackrel{j}{\pi} \hspace{1mm} \in Aut(\stackrel{j}{\mathbb{V}})$ the representation matrix acting on $\stackrel{j}{\mathbb{V}}$.
The representation $j$ of $\g$ on $\stackrel{j}{\mathbb{V}}$ induces a representation of $\gh$ on $\stackrel{j}{\mathbb{V}}$. We will note $\stackrel{j}{\pi}_* \hspace{1mm} \in End(\stackrel{j}{\mathbb{V}})$ the associated representation matrix. 
The representations of $G$ are obtained from those of $\hat{G}$. Namely, they are the representations of $\hat{G}$ that do not see the center, i.e since $\stackrel{j}{\pi}(\pm \mathbb{I})=(\pm 1)^{2j}$, we have $\forall g \in \hat{G},
\stackrel{j}{\pi}(\Lambda(g))=\stackrel{j}{\pi}(g)$ if and only if $j \in \mathbb{N}$.

In order to define a Lagrangian density involving fermions, we need to be able to construct scalar objects with spinors. This implies the definition of a $Spin(3)$-invariant scalar product on $\mathbb{V}$. In pseudo-riemannian spacetime signatures, this bilinear form involves the introduction of the Dirac conjugate $(\psi, \phi) = \overline{\psi} \phi=\psi^{\dagger} \gamma^0 \phi \in \C$, where $^{\dagger}$ denotes Hermitian conjugation on $\mathbb{C}$ and $\gamma^0$ is the gamma matrix (representing the $\gamma$ symbol) corresponding to the timelike direction. It is necessary to introduce the $\gamma^0$ matrix to compensate the transformation property of the gamma matrices under hermitian conjugation $\gamma^{\mu \hspace{.5mm} \dagger}=\gamma^0 \gamma^{\mu} \gamma^0$ in order to construct a real action. However, in the case at hand the gamma matrices are all hermitian and it is sufficient to define the Dirac conjugate by using the $Spin(3)$-invariant hermitian bilinear form on $\mathbb{V}$: $(\psi, \phi) = \overline{\psi} \phi=\psi^{\dagger} \phi \in \C$. Equivalently, one could consider a bilinear form pairing $\V$ with its dual complex conjugate vector space $\overline{\V}^*$.

\subsection{Fermions and spacetime geometry}

We now set up the geometrical structure of the spacetime on which the fermions are evolving.

\subsubsection{Einstein-Cartan geometry}

Let the three-dimensional differential manifold $M$ model our spacetime. Let $FM$ denote the frame bundle over $M$, that is, the principal bundle with base $M$ and structure group $GL(3, \mathbb{R})$ whose typical fiber over each point $p \in M$ consists in the set of all possible basis of the tangent space $T_pM$. Consider a $\mathfrak{gl}(3,\mathbb{R})$-valued linear connection on $FM$. Our spacetime becomes a linearly connected space as tangent vectors over different points of the manifold can be compared. 
One can furthermore endow $M$ with an Euclidean metric $g$, i.e a non-degenerate bilinear form on the tangent space $T_pM$ over each point $p$ of the manifold with purely positive signature, and obtain a metric affine space where the metric and the linear connection are independent geometrical objects. 
Consider now the reduction of $FM$ to the $G=SO(3)$-bundle of orthonormal frames $\mathcal{P}$ characterised by the metric $g$. In the generic case, the linear connection does not know about this reduction as it is independent of the metric.
Locally, for a given section (a frame, i.e an origin and a set of three linearly independent vectors) $e:M \rightarrow \mathcal{P}; p \rightarrow \{e(p)_I\}_I$, $I=1,2,3$, or for a given coordinate basis $\{\frac{\partial}{\partial_{\mu}}\}_{\mu}$, $\mu=1,2,3$, of $T_pM$, the metric is given by

\beq
g=\eta_{IJ} e^I \otimes e^J=g_{\mu \nu} dx^{\mu} \otimes dx^{\nu}.
\eeq 
\\
If we now suppose that the chosen linear connection is compatible with the bundle reduction (i.e that it is reducible) meaning that the restriction of the linear connection to the bundle $\mathcal{P}$ is a principal connection on $\mathcal{P}$, the connection becomes related to the spacetime metric $g$. In that case, the connection obeys the so-called metricity condition stating that the covariant derivative of the metric with respect to the connection vanishes. 
It is equivalent to simply considering a $\mathfrak{so}(3)$-valued connection $\omega$ on $\mathcal{P}$. The
pull-back by a local section of a such connection is a one-form on $M$ taking value in the Lie algebra $\mathfrak{so}(3)$ of $SO(3)$, i.e it is matrix valued. If $\{T_a\}_a$, $a=1,2,3$, denotes a vectorial basis of $\mathfrak{so}(3)$, then in a local chart $(U \subset M, x^{\mu}:U \rightarrow \mathbb{R}^3)$ in the domain of the local section,
 
\beq
\omega=(\omega^I_{\hspace{1mm} J})= -(\omega^J_{\hspace{1mm} I})=\omega_{\mu \hspace{1mm} J}^{\hspace{.5mm} I} dx^{\mu}=\omega^a_{\mu} \otimes T_{a \hspace{1mm} J}^{\hspace{.5mm} I} dx^{\mu}. 
\eeq
\\  
A such connection is called metric and ensures that the lengths and angles are conserved by parallel transport on $M$. The data of a manifold $M$ endowed with a metric and a metric (compatible) connection is called a Riemann-Cartan structure.
We call $D$ the covariant derivative associated to the metric connection. 

The curvature $R=D\omega$ of the connection $\omega$ is given by the matrix valued two-form

\beq
R=(R^I_{\hspace{1mm} J})=D \omega^I_{\hspace{1mm} J} = d \omega^I_{\hspace{1mm} J} + \omega^I_{\hspace{1mm} K} \wedge \omega^K_{\hspace{1mm} J}=\frac{1}{2}R^{\hspace{2mm} I}_{\mu \nu \hspace{.5mm}J} dx^{\mu} \wedge dx^{\nu}.
\eeq
\\
Next, consider the image in $\mathcal{P}$ by a local section of a given point $p \in M$. The result is a basis of $V=\mathbb{R}^3$, i.e a set of three linearly independent vectors $e_I$, orthonormal for the metric $g$. Let $e^I$ denote the associated dual co-frame. It decomposes into a coordinate basis of the cotangent space: $e^I=e_{\mu}^I dx^{\mu}$. The matrix $e_{\mu}^I$ is an isomorphism of vector space between the tangent space over $p$ and $V$ regarded as the vector representation space of $SO(3)$; $\forall v \in T_pM, e^I (v) = v^I \in V$, i.e it expresses any tangent vector into a locally inertial frame where geodesic trajectories are mapped onto straight lines: the triad $e^I$ is the gravitational field. 
We can now define the soldering one form with respect to a given choice of a frame and a coframe $e:TM \rightarrow TM=\mathcal{P} \times_{SO(3)} V$ as $e=e^I \otimes e_I=e_{\mu}^I dx^{\mu}\otimes e_I$.

The torsion $T=De$ of the connection $\omega$ is given by the vector valued two-form

\beq
T^I=D e^I=de^I + \omega^I_{\hspace{1mm} K} \wedge e^K=\frac{1}{2}T^{I}_{\hspace{1.5mm} \mu \nu}dx^{\mu} \wedge dx^{\nu}.
\eeq
\\
In the same way that we have reduced the metric affine structure down to a Riemann-Cartan structure by imposing the metricity condition, we can recover the standard riemannian framework of Einstein's general relativity by restricting the set of metric connections to the torsion free or riemannian connections. In fact, there is a unique metric compatible, torsion free linear connection. It is called the Levi-Civita connection. The Riemann-Cartan geometry \cite{Hehl} that we are considering here is therefore a generalisation of ordinary GR. The idea of extending GR as to include non-zero torsion solutions was proposed by Elie Cartan in the early twenties. He suggested, before the introduction of the modern concept of spin, to relate spacetime torsion to an intrinsic angular momentum of matter. This theory can be obtained as a gauge theory of the Poincar\'e group where torsion and curvature are respectively regarded as the translational and rotational field strengths.
As we have seen, E. Cartan was right: fermions indeed couple to spacetime torsion. 

\subsubsection{Spin structure, Spinor field}

We now wish to introduce fermions on our Riemann-Cartan spacetime. Accordingly , we
equip the manifold $M$ with an extra structure called a spin structure.
A spin structure on $(M,g)$, if it exists \footnote{The existence of a spin structure over a manifold $M$ is related to the vanishing of a given characteristic class called the Stiefel-Whitney class.}, is a principal $\hat{G}$-bundle $\hat{\mathcal{P}}$ with base $M$ and structure group $\hat{G}=Spin(3)$, such that there exists a two-to-one bundle homomorphism $\hat{\chi}$ from $\hat{\mathcal{P}}$ onto the bundle $\mathcal{P}$ of orthonormal frames \footnote{A bundle homomorphism is a map between bundles preserving the fiber structure, i.e commuting with the right action of the structure group on the fiber and compatible with the associated group morphism: here, $\hat{\chi}(p.g)=\hat{\chi}(p) \chi(g)$, $p\in \hat{\mathcal{P}}, g \in \hat{G}$.}. Troughout this paper, we have assumed that a spin structure had been chosen. Note that all principal bundles with compact three dimensional base manifold and compact, connected, simply connected structural group are necessarily trivial. Our spin bundle is therefore trivial and we can consequently choose a global trivialising section once and forall.

The spin bundle over M being given, we can define a spinor field living on our spacetime.
A tangent vector to $M$ can be viewed as an element of the vector bundle associated to the orthonormal frame bundle $\mathcal{P}$ via the fundamental representation of $SO(3)$ on $V=\mathbb{R}^3$. In the same way,    
a spinor field $\psi$ is defined as a section of the vector bundle $SM$ associated to the principal bundle $\hat{\mathcal{P}}$ trough the representation of $Spin(3)$ on $\mathbb{V}$: $SM=\hat{\mathcal{P}} \times_{Spin(3)} \mathbb{V}$. 

To be able to compare the value of the fermionic fields at different points of spacetime, we need a covariant derivative acting on spinors.

\subsubsection{Spin connection}

A spin connection is a principal connection on $\hat{\mathcal{P}}$ such that its image under the homomorphism $\hat{\chi}$ is a metric connection on $\mathcal{P}$.
The explicit form of the spin connection is obtained from the metric connection through the Lie algebra isomorphism $\chi_*$ between $\mathfrak{spin}(3)$ and $\mathfrak{so}(3)$. By using the defining relation \eqref{spin} of $Spin(3)$ at the level of the Lie algebras (i.e by differentiation), we obtain a unique correspondence between the generators of $\mathfrak{spin}(3)$ and those of $\mathfrak{so}(3)$. In a chart with local coordinates $x^{\mu}$, the components of the $\mathfrak{spin}(3)$-valued connection one-form $\hat{\omega}$ are given in terms of the components of the metric connection :

\beq
\hat{\omega}_{\mu} = -\frac{1}{4} \omega_{\mu}^{IJ} \gamma_I \gamma_J, 
\eeq
\\
where indices are raised and lowered with the metric $\eta$.
In the representation $\rho_*$ of $\mathfrak{spin}(3)$ induced by the representation $\rho$ of the Clifford algebra $\mathcal{C}(3)$ in terms of the Pauli algebra, the spin connection is given by
$\rho_*(\hat{\omega}_{\mu}) = -\frac{1}{4} \omega_{\mu}^{IJ} \sigma_I \sigma_J$
 
We can now define the covariant derivative one form $\nabla \psi=\nabla_{\mu} \psi dx^{\mu}$ of a spinor field $\psi \in \Gamma(SM)$ in the spin connection defined above
\beq
\nabla_{\mu} \psi = \partial_{\mu} \psi + \rho_*(\hat{\omega}_{\mu}) \psi.
\eeq
\\

We also define a covariant derivative for the Dirac conjugate, or co-spinor, $\overline{\psi}$:

\beq
\nabla_{\mu} \overline{\psi} = \partial_{\mu} \overline{\psi} - \overline{\psi} \rho_*(\hat{\omega}_{\mu}).
\eeq
\\

\end{document}